\documentclass[letterpaper, 11pt]{article}

\usepackage{jheppub}
\usepackage{bm}
\usepackage{graphicx}
\usepackage{epstopdf}
\usepackage{amsmath, amssymb}

\usepackage{amsfonts,amssymb,epsfig,amsmath,mathtools,tabu}
\usepackage{verbatim}

\usepackage[inline]{enumitem}

\usepackage{hyperref,caption,subcaption}
\usepackage{xcolor,tikz,graphicx,afterpage}
\usetikzlibrary{shapes.geometric,positioning}

\newcommand{\be}{\begin{eqnarray}}
\newcommand{\ee}{\end{eqnarray}}
\newcommand{\nn}{\nonumber}
\newcommand{\bn}{\begin{enumerate}}
\newcommand{\en}{\end{enumerate}}

\def\IR{\mathbb{R}}

\def\CF{{\cal F}}

\def\CH{{\cal H}}
\def\CI{{\cal I}}

\def\CN{{\cal N}}
\def\CO{{\cal O}}

\def\CQ{{\cal Q}}

\def\CS{{\cal S}}

\def\CX{{\cal X}}

\def\CZ{{\cal Z}}

\def\a{\alpha}
\def\b{\beta}

\def\m{\mu}

\def\w{\omega}
\def\half{\frac{1}{2}}

\def\Tr{{\rm Tr}}

\title{A 4d $\CN=1$ Cardy Formula}

\author[a, e]{Joonho Kim,}
\author[b]{Seok Kim}
\author[a, c, d]{and Jaewon Song}
\affiliation[a]{School of Physics, Korea Institute for Advanced Study, Seoul 02455, Korea}
\affiliation[b]{Department of Physics and Astronomy \& Center for Theoretical Physics\\ Seoul National University, Seoul 08826, Korea}
\affiliation[c]{Asia Pacific Center for Theoretical Physics, Pohang, Gyeongbuk 37673, Korea}
\affiliation[d]{Department of Physics, Korea Advanced Institute of Science and Technology, Daejeon 34141, Korea}
\affiliation[e]{Institute for Advanced Study, Princeton, NJ 08540, USA.}
\emailAdd{joonhokim@ias.edu}
\emailAdd{skim@phya.snu.ac.kr}
\emailAdd{jaewon.song@kaist.ac.kr}

\abstract{
We study the asymptotic behavior of the (modified) superconformal index for 4d $\CN=1$ gauge theory. By considering complexified chemical potential, we find that the `high-temperature limit' of the index can be written in terms of the conformal anomalies $3c-2a$. 
We also find macroscopic entropy from our asymptotic free energy when the Hofman-Maldacena bound $1/2 < a/c < 3/2$ for the interacting SCFT is satisfied. 
We study $\CN=1$ theories that are dual to AdS$_5 \times Y^{p, p}$ and find that the Cardy limit of our index accounts for the Bekenstein-Hawking entropy of large black holes. 
}

\preprint{KIAS-P19015, SNUTP19-002}

\begin{document}
\maketitle

\section{Introduction}
The four-dimensional supersymmetric index (or the superconformal index) counts certain protected states on a three-sphere \cite{Kinney:2005ej, Romelsberger:2005eg}, which can be written as
\begin{align} \label{eq:sci}
 \CI( \w_1, \w_2) = \Tr_{\CH} (-1)^F e^{-\w_1 (J_1+R/2)} e^{- \w_2 (J_2+R/2)} = \Tr_{\CH} (-1)^F p^{J_1+R/2} q^{J_2+R/2}  \ ,
\end{align}
where the trace is over the states on $S^3$. Here $J_{1, 2}, R$ are the generators of the angular momenta and $R$-charge. The $\w_{1, 2}$ are the chemical potentials for the angular momenta shifted by $R$-charge. We also use the fugacities $p \equiv e^{-\w_1}, q \equiv e^{-\w_2}$ frequently throughout the paper. 
Since the supersymmetric index is invariant under the renormalization group flow and also under the change of marginal couplings, it can often be computed using the weak-coupling limit of the gauge theory. Hence it provides a useful tool to investigate the non-perturbative aspects of superconformal theories. See the review \cite{Rastelli:2016tbz} and the references therein  for more details.

We call the limit $|\w_i | \ll 1$ as the Cardy limit in analogy with the high-temperature limit of the Cardy's formula for two-dimensional CFT \cite{Cardy:1986ie}. In our case, the `Cardy limit' actually corresponds to the large charge (or angular momentum) limit at zero temperature since the index counts BPS states. The Cardy limit of the 4d superconformal index has been studied by di Pietro and Komargodski \cite{DiPietro:2014bca}. They found that the asymptotic behavior of the index \eqref{eq:sci} can be written in terms of the conformal anomalies as
\begin{align}
 \CI(\w) \sim \exp \left( \frac{16\pi^2}{3 \w} (c-a) \right) \ , 
\end{align}
in the $\w_1 = \w_2 = \w \ll 1$ limit assuming the chemical potentials are strictly real. It was noticed by \cite{Ardehali:2015bla} that this formula can receive corrections for the theories having $a>c$.
One of the original motivation to introduce the superconformal index is to reproduce the entropy of the supersymmetric black holes in AdS$_5 \times S^5$ \cite{Gutowski:2004ez, Gutowski:2004yv, Chong:2005hr, Kunduri:2006ek}. It has been believed for more than a decade that this goal cannot be achieved from the index due to severe boson-fermion cancellations. 
This is also consistent with the di Pietro-Komargodski's formula for $a \approx c$ since the free energy with $(-1)^F$ seems to be much smaller than the degrees of freedom counted by the central charges. 

However, it was recently realized that the index of $\CN=4$ $U(N)$ SYM theory does capture the black hole entropy once we allow the chemical potentials to have imaginary parts \cite{Cabo-Bizet:2018ehj, Choi:2018hmj, Choi:2018vbz, Benini:2018ywd}.  In particular, it was shown in \cite{Choi:2018hmj} that when the chemical potentials for the angular momentum $|\w_{1,2}|$ are taken to be small, $\textrm{log}(\CI )$ scales as $N^2$ and reproduces the `black hole entropy function' of \cite{Hosseini:2017mds} in the Cardy regime. 
See also \cite{ArabiArdehali:2019tdm,Honda:2019cio}.

In this paper, we generalize the analysis of \cite{Choi:2018hmj} to arbitrary $\CN=1$ superconformal theories with finite and general central charges. One obvious difference between generic $\CN=1$ theory and $\CN=4$ theory is that the two central charges $(a, c)$ are not necessarily equal. 
In addition, one crucial difference is that the superconformal $R$-charges for the $\CN=1$ theories are not quantized and can be any real numbers.\footnote{It is widely believed that the $R$-charges for the chiral operators in an $\CN=1$ theory are algebraic numbers \cite{Intriligator:2003jj}, whereas for $\CN=2$ are rational \cite{Caorsi:2018zsq, Argyres:2018urp}. For $\CN \ge 3$, $R$-charges are integers up on suitable normalization.} Therefore, a straight-forward generalization of \cite{Choi:2018hmj} requires a slight twist. We define a modified version of the superconformal index as
\begin{align} \label{eq:Ridx}
 \CI(\w_1, \w_2) = \Tr_{\CH} \left[ e^{\pi i R} p^{J_1 + R/2} q^{J_2 + R/2} \right] \ , 
\end{align}
where $(-1)^F$ is replaced by $e^{\pi i R}$. This form indeed qualifies as a Witten index since the supercharge shifts $R$ by exactly one unit.\footnote{A similar definition of the index was used in \cite{Cordova:2016uwk} in the case of $\CN=2$ theory.} This form of the index contains equivalent information as the conventional one since we simply multiplied an extra phase factor for each contribution coming from short multiplets. 
However, it turns out that the modified index in the Cardy limit captures more entropy than the ordinary index. It enables us to extract enough amount of degrees of freedom that accounts for the black hole entropy in the large $N$ limit. 
For the models with flavor symmetries, it often happens that shifting their chemical potentials by suitable imaginary amounts changing $e^{ \pi i R }$ into the ordinary $(-1)^F$. In these cases, using \eqref{eq:Ridx} is equivalent to turning on the complex flavor chemical potentials for the ordinary index, which allows us to see the true large charge saddle points.

We find that the modified index in the Cardy limit is given as 
\begin{align} \label{eq:Fformula}
 \textrm{log} \CI(\w_1, \w_2) \sim \frac{8\Delta^3}{27 \w_1 \w_2 }(5a-3c) +\frac{8\pi^2 \Delta}{3 \w_1 \w_2}  (a-c) \ , 
\end{align}
with $\Delta = \frac{\w_1 + \w_2}{2} - \pi i$. Here $\textrm{Re}(\Delta) > 0$ and $\textrm{Re}(\w_{1, 2})>0$. Once we allow the chemical potential for the angular momentum to have both real and imaginary part, the index in the Cardy limit $|\w_{1, 2}| \ll 1$ can be written as (with $\w_1 = \w_2 = \w \equiv \w_R + i \w_I$)
\begin{align}
 \textrm{Re} [\textrm{log}\CI(\w)] \sim \frac{32 \pi^3 \w_I \w_R }{27 |\w|^4} (3c-2a) \ . 
\end{align}
This is our Cardy formula. 
For an interacting unitary SCFT, $(3c-2a)$ is always positive \cite{Hofman:2008ar} (it is zero for a free vector multiplet). Therefore the real part of the index always grows exponentially once we have positive $\w_I$. 

The index can be written as 
\begin{align}
 \CI(\w_1, \w_2) = \sum_{\textrm{BPS states}} \Omega(j_1, j_2) e^{-\w_1 j_1 - \w_2 j_2} \ , 
\end{align}
where $\Omega(j_1, j_2)$ gives the lower bound on degeneracies of the BPS states with given charges. In principle, one can perform an inverse Laplace transformation to extract $\Omega(j_1, j_2)$. In the Cardy limit, one can simplify the procedure via saddle point approximation. In the end, it is equivalent to performing the Legendre transformation on the Cardy free energy $\textrm{log}(\CI)$.
We perform the Legendre transformation of the Cardy free energy to obtain the entropy of a microcanonical ensemble of the states with large angular momentum. It turns out that the most dominant saddle point is complex-valued with a positive imaginary part $\w_I>0$. At the saddle point, the entropy becomes
\begin{align}
 \textrm{Re}(S) = \textrm{log}\Omega(J) \sim (3c-2a)^{1/3} J^{2/3} \ , 
\end{align}
for $J_1=J_2=J$ so that it is positive whenever $a/c < 3/2$, which is identical to the one side of the Hofman-Maldacena bound \cite{Hofman:2008ar}. 

For a holographic theory (such as $\CN=4$ SYM) with $a \approx c$, we easily see that the free energy given as above scales as $a \sim N^2$. Our formula distinctively differs from the earlier work of \cite{DiPietro:2014bca} where $a-c$ played the role of proportionality constant. In our case, we get $\frac{3c-2a}{\omega^2}$ as the leading asymptotic behavior instead of $\frac{c-a}{\omega}$. 
We also see that in $a=c$ limit, \eqref{eq:Fformula} reduces to the so-called the entropy function for the AdS$_5$ black hole. It was advocated in \cite{Hosseini:2017mds, Cabo-Bizet:2018ehj} that Casimir energy accounts for the black hole entropy. This is reminiscent of the two-dimensional Cardy formula, where the Casimir energy (which is the low-temperature behavior fixed by the central charge) is indeed related to the high-temperature asymptotics. Therefore it is quite possible that some version of modular invariance is hidden in this setup. 

We also apply our result to the case with $\CN=1$ SCFT dual to type IIB string theory on $AdS_5 \times Y^{p, q}$ \cite{Benvenuti:2004dy}. We find that when $p=q$, the Cardy free energy can be written in the same form as that of the $\CN=4$ SYM theory upon mapping the chemical potentials. This allows us to perform Legendre transformation to obtain the macroscopic entropy of the supersymmetric black holes of \cite{Gutowski:2004ez, Gutowski:2004yv, Kunduri:2006ek}. 

The outline of this paper is as follows. In Section \ref{sec:index}, we give a detailed analysis of the (modified) superconformal index in our Cardy limit. In Section \ref{sec:eff-action}, we derive the same formula using the background field analysis on three-sphere. In Section \ref{sec:index}, we focus on Lagrangian gauge theories. However, with our intrinsic anomaly-based analysis of Section \ref{sec:eff-action}, we expect that the result is true for non-Lagrangian theories as well. In Section \ref{sec:examples}, we consider various examples to demonstrate that the dominant saddle point in the holonomy integral is at the origin which preserves the full gauge symmetry. Then in Section \ref{sec:entropy}, we perform the Legendre transformation of the asymptotic free energy to obtain the asymptotic entropy of the microcanonical ensemble of fixed charges. 


\section{Asymptotic index of $\mathcal{N}=1$ gauge theory} \label{sec:index}

Let us consider the general partition function of $\mathcal{N}=1$ gauge theory on $S^3 \times \mathbf{R}$,
\begin{align} \label{eq:Zdef}
 \CZ(\b, \w_1, \w_2, \Delta) = \Tr \left[ e^{-\b E - \w_1 J_1 - \w_2 J_2 - \Delta R + i \mathbf{x}\cdot\mathbf{f}} \right] = \Tr \left[ e^{-\beta E} p^{J_1} q^{J_2} t^{R} \prod_{a} z_a^{f_a} \right] \ , 
\end{align}
where $(E, J_1, J_2, R)$ are the Cartans of the superconformal algebra $\mathfrak{su}(2, 2|1)$ and $
p = e^{-\w_1}, q = e^{-\w_2},  t= e^{-\Delta}, z_a = e^{i x_a}$. Here we choose the chemical potentials $\w_{1,2}$ as the conjugate to the angular momentum associated to two $\IR^2$ planes inside $\IR^4$. We define $J_1 = j_L+j_R, J_2 = j_L - j_R$ with $(j_L, j_R)$ being the generators of the Lorentz group $SU(2)_L \times SU(2)_R \subset SO(4)$. 
When there is a flavor symmetry in the theory, we introduce the flavor chemical potentials $\mathbf{x} = (x_1, \cdots, x_{|F|})$, being conjugate to the Cartan generators $\mathbf{f} = (f_1, \cdots, f_{|F|})$ of the flavor symmetry group $F$.

The $\mathcal{N}=1$ supercharges are $\{\mathcal{Q}_\alpha, \mathcal{Q}^{\dagger \alpha},  \tilde{\mathcal{Q}}_{\dot{\alpha}}, \tilde{\mathcal{Q}}^{\dagger\dot{\alpha}}\}$, where $\alpha$ and $\dot{\alpha}$ denote doublet indices for $SU(2)_L$ and $SU(2)_R$, respectively. Here we assign their $R$-charges to be $[R,\mathcal{Q}_{\alpha}]= +\mathcal{Q}_{\alpha}$ and $[R,\tilde{\mathcal{Q}}_{\dot\alpha}]= -\tilde{\mathcal{Q}}_{\dot\alpha}$. For the following pairs of supercharges, their anti-commutation relations are given by
\begin{align}
\begin{split}
    \delta_\pm = \{\mathcal{Q}_\pm, \mathcal{Q}^{\dagger\pm}\} &= E \pm 2j_L - \frac{3}{2}R \ , \\
    \tilde{\delta}_{\dot \pm} = \{\tilde{\mathcal{Q}}_{\dot \pm}, \tilde{\mathcal{Q}}^{\dagger\dot \pm}\} &= E \pm 2j_R + \frac{3}{2}R \ .
\end{split}
\end{align}
When the supercharges are acted by the operators inside the trace formula \eqref{eq:Zdef}, they transform as
\begin{align}
\begin{split}
 e^{-\Delta R - \w_1 J_1 - \w_2 J_2} \CQ_{\pm} &= e^{\frac{- 2\Delta \mp (\w_1 + \w_2)}{2}}\CQ_{\pm} e^{-\Delta R - \w_1 J_1 - \w_2 J_2} \ , \\
 e^{-\Delta R - \w_1 J_1 - \w_2 J_2} \tilde{\CQ}_{\dot{\pm}} &= e^{\frac{2\Delta \mp (\w_1 - \w_2)}{2}}\tilde{\CQ}_{\pm} e^{-\Delta R - \w_1 J_1 - \w_2 J_2} \ . 
\end{split}
\end{align}
Once we take the limit $\beta \rightarrow 0$ with the following constraint between the chemical potentials, the above partition function becomes the Witten index (or the superconformal index) preserving the corresponding supercharges:
\begin{align} \label{eq:constraint}
\begin{array}{ccc}
 \CQ_+ : & 2\Delta + \w_1 + \w_2 = - 2\pi i &~~(\textrm{mod }4\pi i) \\
 \CQ_- : &  2\Delta - \w_1 - \w_2 = - 2\pi i &~~(\textrm{mod }4\pi i) \\
 \tilde{\CQ}_{\dot{+}} : & 2\Delta - \w_1 + \w_2  = - 2\pi i &~~(\textrm{mod }4\pi i) \\
 \tilde{\CQ}_{\dot{-}} : & 2\Delta + \w_1 - \w_2  = - 2\pi i &~~(\textrm{mod }4\pi i) 
\end{array}
\end{align}
Then the index receives contributions only from the states with either $\delta_\pm=0$ or $\tilde{\delta}_\pm  =0$ depending on the choice of the constraint. 
For example, if we choose $2\Delta - \w_1 - \w_2 = - 2 \pi i$, we set $t^2 = p q$ with the $e^{\pi i R}$ insertion inside the trace. It gives the index $\CI_-$ which is the trace over the states with $\delta_-=0$, i.e.,
\begin{align} \label{eq:index}
 \CZ(\beta, p, q, t) \to \CI_{-} (p, q) = \Tr_{\delta_-=0} \left[ e^{\pi i R} p^{J_1 + \frac{R}{2}} q^{J_2 + \frac{R}{2}} \right] \ . 
\end{align}
This is the familiar definition of the $\CN=1$ superconformal index in the literature except that $(-1)^F$ is replaced by $e^{\pi i R}$. It is effectively the same as the insertion of $(-1)^F$ since the supercharges will map states with $R$ to $(R\pm 1)$. The main difference is that we dress each supermultiplet by the phase determined by the $R$-charge of the top component. 

Another thing to note is that the choice of the supercharge will set the signs for (the real part of) the chemical potentials. The current choice ($\CQ_-$) sets the real part of $\Delta$ and $\w_+ \equiv \frac{\w_1 + \w_2}{2}$ to be of the same sign. The index $\CI_-$ gets contributions from the states with $j_L+R/2 \ge 0$. To see this, note that the index $\CI_-$ is obtained from the states with $\delta_-=0$ but also any states in the theory satisfies $\delta_+ \ge 0, \tilde{\delta}_{\dot{\pm}}  \ge 0$. We can easily see that $j_L+R/2 =( 2\delta_+ +\tilde{\delta}_{\dot{+}} + \tilde{\delta}_{\dot{-}})/12 \ge 0$. Therefore the index is convergent as a series expansion in terms of the fugacity $(p q)$. Other choices of supercharges work with different sign choices. We proceed with the index $\CI_-$ in the rest of this paper. 

\subsection{Letter partition function}
Here we study the partition function \eqref{eq:Zdef} for a very weakly coupled gauge theory as was done in the case of $\CN=4$ SYM in \cite{Aharony:2003sx}. In the end, we will study the index that can be evaluated reliably at strong-coupling by imposing the supersymmetry condition \eqref{eq:constraint}. But this turns out to be a useful exercise to correctly identify the contributing factors to the index and the choice of signs for the chemical potentials. 

Let us first evaluate the single letter partition function $f^V$ for an $\mathcal{N}=1$ vector multiplet. The operators $F_{\mu\nu}$ and $(\lambda_{\alpha}, \lambda_{\dot{\alpha}})$ are subject to the equations of motion, i.e., $\partial^\mu F_{\mu\nu} = 0$ and $\Gamma^{\mu \alpha\dot{\alpha}} \partial_{\mu}\lambda_{\alpha} = \Gamma^{\mu \alpha\dot{\alpha}} \partial_{\mu}\lambda_{\dot\alpha}  = 0$.
Thus $f^V$ is obtained by adding up the contributions from the component fields, minus that of the equations of motion. Each part in an $\mathcal{N}=1$ vector multiplet contributing to the partition function is summarized in Table~\ref{tbl:vec}. We also write the index $\CI_-$ which can be obtained in $\beta \to 0$ limit with $t^2 = pq$ and insertion of $e^{\pi i R}$. 
\begin{table}[h]
    \centering
    $\def\arraystretch{1.3}
    \begin{array}{c||c|c|c||c | c }
        \text{Letter} & SU(2)_L \times SU(2)_R & E & R & f^V & \CI_- \\ 
        \hline
        F_{\mu\nu}  & (\mathbf{3}, \mathbf{1}) \oplus (\mathbf{1}, \mathbf{3}) & 2 & 0 & e^{-2\beta}\, (\chi_3(pq) + \chi_3 (p/q)) & pq \\
        \lambda_{\alpha}  & (\mathbf{2}, \mathbf{1}) &  3/2 & -1 & e^{-3\beta/2}\, t^{-1}\, \chi_2(p q) & 0 \\
        \bar{\lambda}_{\dot\alpha}  & (\mathbf{1}, \mathbf{2}) &  3/2 & +1 & e^{-3\beta/2}\, t^{1}\, \chi_2(p/q) & -p-q \\ 
        \hline
        \partial^\mu F_{\mu\nu} & 2 (\mathbf{2}, \mathbf{2}) &  3 & 0 & - e^{-3\beta}\, \chi_2(pq)\chi_2 (p/q) & 0 \\
        \Gamma^{\mu \alpha\dot{\alpha}} \partial_{\mu}\lambda_{\alpha}  & (\mathbf{1}, \mathbf{2}) &  5/2 & -1 & - e^{-5\beta/2}\, t^{-1}\, \chi_2(p/q) & 0 \\
        \Gamma^{\mu \alpha\dot{\alpha}} \partial_{\mu}\bar{\lambda}_{\dot\alpha}  & (\mathbf{2}, \mathbf{1}) &  5/2 & +1 & - e^{-5\beta/2}\, t^{+1}\, \chi_2(pq) & pq \\ \hline
        \partial^\mu\partial^\nu F_{\mu\nu} & 2\,(\mathbf{1}) &  4 & 0 & e^{-4\beta} & 0 \\
        \hline
        \partial_{\pm \dot{\pm} } & (\mathbf{2}, \mathbf{2}) & 1 & 0 &  e^{-\beta}p^{\pm} q^{\pm} & p, q
    \end{array}$
    \caption{Letter partition function of an $\mathcal{N}=1$ vector multiplet }
    \label{tbl:vec}\vspace{0.1cm}
\end{table}
In the table, $\chi_d(z)$ denotes the character for the $d$-dimensional irreducible representation of $SU(2)$ given as $\chi_2(z) = z^{\half} + z^{-\half}$ and $\chi_3(z)=z+1+z^{-1}$. 

Taking into account the tower of derivatives, $(\partial)^n$, acting on the letters and the equations of motion, the bosonic operator $F_{\mu\nu}$ contributes to 
\begin{align}
    f_B^V = \frac{e^{-2\beta} \left(\chi_3(pq) + \chi_3(p/q) \right) - 2e^{-3\beta} \chi_2(pq)\chi_2(p/q) + 2 e^{-4\beta}}
    {(1-e^{-\beta}p)(1-e^{-\beta}p^{-1})(1-e^{-\beta}q)(1-e^{-\beta}q^{-1})} \cdot \chi_{\bf adj}^G \ , 
\end{align}
where $G$ refers to the gauge group and $\chi_{\mathbf{R}}^G$ refers to the character for the representation $\mathbf{R}$ for the gauge group $G$. 
The last term in the numerator was added to compensate for subtraction of $\partial^\mu\partial^\nu F_{\mu\nu}$ which vanishes identically. 
Similarly, the fermionic operator $(\lambda_{\alpha}, \lambda_{\dot\alpha})$ contributes to
\begin{align}
    f_F^V &= \frac{e^{-\frac{3}{2}\beta}  \chi_2(pq) t^{-1} - e^{-\frac{5}{2}\beta}\chi_2(p/q) t^{-1} + e^{-\frac{3}{2}\beta} \chi_2(p/q)t^{1}  - e^{-\frac{5}{2}\beta}\chi_2(pq)t^{1} }
    {(1-e^{-\beta}p)(1-e^{-\beta}p^{-1})(1-e^{-\beta}q)(1-e^{-\beta}q^{-1})} \cdot \chi_{\bf adj}^G \ .
\end{align}

Now let us consider the single letter partition function $f^\mathcal{X}$ for a chiral multiplet $\mathcal{X}$ in the representation $(\mathbf{R}, \mathbf{F})$ of $G \times F$. 
Following the Romelsberger's prescription \cite{Romelsberger:2007ec}, we do not fix the $U(1)_R$-charge of the chiral multiplet to be that of the free field $2/3$.\footnote{In our convention, this is the charge of the scalar in the anti-chiral multiplet. We will always call the $R$-charge of a chiral multiplet as that of $\Phi^\dagger$ in Table \ref{tbl:chi}.} Instead, we leave it as a free parameter $r_{\CX}$ that will be fixed for an interacting theory using anomaly cancellation or $a$-maximization \cite{Intriligator:2003jj}. The letters in a chiral multiplet and their contribution to the partition functions/indices are summarized in Table~\ref{tbl:chi}.
\begin{table}[h]
    \centering
    $\def\arraystretch{1.3}
    \begin{array}{c||c|c|c||c|c}
        \text{Letter} & SU(2)_L \times SU(2)_R & E & R & f^{\CX} & \CI_- \\ 
        \hline
        \Phi  & (\mathbf{1}, \mathbf{1}) &  \frac{3}{2}r_{\mathcal{X}} & -r_{\mathcal{X}} & e^{-(3r_{\mathcal{X}}/2)\cdot \beta}\,t^{-r_{\mathcal{X}}} & 0 \\
        \Phi^\dagger  & (\mathbf{1}, \mathbf{1}) &  \frac{3}{2}r_{\mathcal{X}} & r_{\mathcal{X}} & e^{-(3r_{\mathcal{X}}/2)\cdot \beta}\, t^{-r_{\mathcal{X}}} & (pq)^{\frac{r_\chi}{2}}\\
        \psi_{\alpha}  & (\mathbf{2}, \mathbf{1}) &  \frac{3}{2}r_{\mathcal{X}} + \frac{1}{2} &  - r_{\mathcal{X}} + 1 & e^{-(3r_{\mathcal{X}}+1)/2\cdot \beta}\, t^{-r_{\mathcal{X}}+1}\, \chi_2(pq)& -(-pq)^{1 - \frac{r_\chi}{2}} \\
        \bar{\psi}_{\dot\alpha}  & (\mathbf{1}, \mathbf{2}) &  \frac{3}{2}r_{\mathcal{X}} + \frac{1}{2} &  r_{\mathcal{X}}-1 & e^{-(3r_{\mathcal{X}}+1)/2\cdot \beta}\, t^{r_{\mathcal{X}}-1}\, \chi_2(p/q) & 0 \\ 
        \hline
        \partial^2\Phi  & (\mathbf{1}, \mathbf{1}) &  \frac{3}{2}r_{\mathcal{X}}+2 & -r_{\mathcal{X}} & - e^{-(3r_{\mathcal{X}}/2+2)\cdot \beta}\,t^{-r_{\mathcal{X}}}& 0 \\
        \partial^2\Phi^\dagger  & (\mathbf{1}, \mathbf{1}) &  \frac{3}{2}r_{\mathcal{X}}+2 & r_{\mathcal{X}}  &-  e^{-(3r_{\mathcal{X}}/2+2)\cdot \beta}\,t^{r_{\mathcal{X}}}& 0\\
        \Gamma^{\mu \alpha\dot{\alpha}} \partial_{\mu}\psi_{\alpha}  & (\mathbf{1}, \mathbf{2} ) & \frac{3}{2}r_{\mathcal{X}}+\frac{3}{2} & -r_{\mathcal{X}} + 1 &-  e^{-(3r_{\mathcal{X}}+3)/2\cdot \beta}\, t^{-r_{\mathcal{X}}+1}\, \chi_2(p/q) & 0 \\ 
        \Gamma^{\mu \alpha\dot{\alpha}} \partial_{\mu}\bar{\psi}_{\dot\alpha}  & (\mathbf{2}, \mathbf{1}) & \frac{3}{2}r_{\mathcal{X}}+\frac{3}{2} & r_{\mathcal{X}}-1 &-  e^{-(3r_{\mathcal{X}}+3)/2\cdot \beta}\, t^{r_{\mathcal{X}}-1}\, \chi_2(pq) & 0 \\
        \hline
        \partial_{\pm \dot{\pm} } & (\mathbf{2}, \mathbf{2}) & 1 & 0 &  e^{-\beta}p^{\pm} q^{\pm} & p, q
    \end{array}$
    \caption{Letter partition function of an $\mathcal{N}=1$ chiral multiplet}
    \label{tbl:chi}\vspace{0.1cm}
\end{table}    
 The scalar fields $(\Phi, \Phi^\dagger)$ are subject to $\partial^2 \Phi  = \partial^2 \Phi^\dagger = 0$. The fermionic fields $(\psi_{\alpha}, \psi_{\dot\alpha})$ satisfy the Dirac equation $\Gamma^{\mu \alpha\dot{\alpha}} \partial_\mu \psi_{\alpha} = \Gamma^{\mu \alpha\dot{\alpha}} \partial_\mu \psi_{\dot\alpha} = 0$. Combining the contributions from each letter, we obtain 
\begin{align}
    f_B^{\mathcal{X}} &= \frac{e^{-(3r_{\mathcal{X}}/2) \beta}(1-e^{-2\beta}) \left(t^{-r_{\mathcal{X}}}\,   \chi_{\bf{R}}^G\,\chi_{\bf{F}}^F  + t^{r_{\mathcal{X}}} \,  \chi_{\overline{\bf{R}}}^G \,\chi_{\overline{\bf{F}}}^F\right)}
    {(1-e^{-\beta}p)(1-e^{-\beta}p^{-1})(1-e^{-\beta}q)(1-e^{-\beta}q^{-1})} \ , \\
    \begin{split}
    f_F^{\mathcal{X}} &= \frac{e^{-(3r_{\mathcal{X}} + 1)/2\cdot \beta } t^{-r_{\mathcal{X}}+1} \left( \chi_2(pq) - e^{- \beta } \,\chi_2(p/q)\right)  \chi_{\bf{R}}^G\chi_{\bf{F}}^F}
    {(1-e^{-\beta}p)(1-e^{-\beta}p^{-1})(1-e^{-\beta}q)(1-e^{-\beta}q^{-1})}\\
    &\quad+ \frac{e^{-(3r_{\mathcal{X}} + 1)/2\cdot \beta } t^{r_{\mathcal{X}}-1}  \left( \chi_2(p/q) - e^{- \beta } \,\chi_2(pq) \right)  \chi_{\overline{\bf{R}}}^G\chi_{\overline{\bf{F}}}^F}
    {(1-e^{-\beta}p)(1-e^{-\beta}p^{-1})(1-e^{-\beta}q)(1-e^{-\beta}q^{-1})} \ .
    \end{split}
\end{align}

Once we are given the single-letter partition function, the (gauge-variant) multi-letter partition function is obtained by taking the Plethystic exponential (PE) as 
\begin{align}
    \label{eq:part}
    \CZ = \exp{\left[\sum_{m=1}^\infty \frac{1}{m} \sum_{\varphi} \left(f_B^{\varphi}(\cdot^m) + (-1)^{m+1} f_F^{\varphi}(\cdot^m)\right)\right]} \ , 
\end{align}
where $(\cdot^m)$ indicates that all chemical potentials multiplied by $m$, including the ones for the gauge symmetry. Here we distinguish the bosonic and fermionic PE to take into account the spin-statistics. The $\varphi$-summation runs over all the $\mathcal{N}=1$ multiplets in a given QFT. To obtain the gauge-invariant partition function we integrate $\CZ$ over the gauge group with Haar measure.

\subsection{Cardy limit of the superconformal index}
\label{subsec:asymptotic-index}
Let us consider a generic $\CN=1$ gauge theory with gauge group $G$ and a number of chiral multiplets $\{\CX \}$ under the representations $\mathbf{R}_{\CX}$ with $R$-charges $r_{\CX}$. The central charges for $\CN=1$ SCFT are given in terms of trace anomalies of the superconformal $R$-current as \cite{Anselmi:1997am}
\begin{align}
 a = \frac{3}{32} (3 \Tr R^3 - \Tr R) \ , \quad c = \frac{1}{32} ( 9 \Tr R^3 - 5 \Tr R) \ . 
\end{align}
For the gauge theory at our hand, we find
\begin{align}
 \Tr R =  |G| + \sum_{\CX} (r_{\CX} -1) |r_{\CX} | \ , \quad
 \Tr R^3 =  |G| + \sum_{\CX} (r_{\CX}-1)^3 |r_{\CX}| \ , 
\end{align}
where the sum is over all the chiral multiplets in the theory. Here the first terms come from the gauginos and the second terms from the fermions in the chiral multiplets. 
Combining the two expressions, we get
\begin{align}
 a &= \frac{3}{32} \left( 2|G| + \sum_{\CX} |\mathbf{R}_{\CX}| \left( 3(r_{\CX}-1)^3 - (r_{\CX}-1) \right) \right)  \ , \\
 c &= \frac{1}{32} \left(  4|G| + \sum_{\CX} |\mathbf{R}_{\CX}| \left( 9(r_{\CX}-1)^3 - 5 (r_{\CX}-1) \right) \right) \ . 
\end{align}
The superconformal $R$-charges $r_{\CX}$ are constrained via gauge anomaly cancellation
\begin{align}
 \Tr R G G = 0 \quad \Leftrightarrow \quad d(G) + \sum_i d(\mathbf{R}_{\CX}) (r_{\CX} - 1) = 0 \ , 
\end{align}
where $d(\mathbf{R})$ refers to the Dynkin index of the representation $\mathbf{R}$. Sometimes, anomaly-free condition is not enough to fix the $R$-charge. In this case, one can use the $a$-maximization procedure \cite{Intriligator:2003jj} to fix the $R$-charges. We aim to express the asymptotic expression for the partition function (index) in terms of the central charges. 

Now, let us study the asymptotic behavior of the superconformal index in the Cardy limit, i.e., $|\omega_1|, |\omega_2 | \ll 1$. 
The superconformal index for a gauge theory is obtained by the gauge invariant projection of the letter index \eqref{eq:part}. This is done by integrating over the Haar measure (also referred to as the Molien integral) as
\begin{align}
    \label{eq:sci-integral}
    \mathcal{I} = 
    \left. \int \prod_{i=1}^{|G|} d\alpha_i \ \prod_{\mathbf{\rho} \in \Delta_G^+} 2\sin^2 \left(\frac{\mathbf{\rho}\cdot \mathbf{\alpha}}{2}\right) \exp \left[ \sum_{m=1}^{\infty} \sum_\varphi \frac{f^\varphi_B(\cdot^m) + (-1)^{m+1} f^\varphi_F(\cdot^m) }{m} \right] \right|_{\eqref{eq:constraint}} \ , 
\end{align}
where $\Delta_G^+$ refers to the set of all positive roots of $G$. 
The integrand in our Cardy limit can be greatly simplified. 
First, let us take $\beta \to 0$ with the index constraint $2 \Delta - \w_1 - \w_2 = -2\pi i $ imposed. For a vector multiplet, we get
\begin{align}
    \CZ_V \Big|_{\eqref{eq:constraint}} =  \exp{\left[\sum_{n\ge 1} \frac{1}{n}\left(1 - \frac{(-1)^{n}\, 2\sinh(n\Delta)}{2\sinh(n\omega_1/2)\,2\sinh(n\omega_2/2)}\right)\cdot\chi_\text{\bf adj}(n\alpha)  \right]} \ .
\end{align}
Here the $\sum_n 1/n$ inside the exponential comes from the bosonic part, and the rest comes from the fermionic part. 

The term $\sum_n \big(\frac{1}{n}\cdot \chi_\text{\bf adj}(n\alpha) \big)$ in the exponent mostly cancels with the Haar measure. The Haar measure can be written as
\begin{align}
 \frac{1}{|W_G|} \prod_{\rho \in \Delta_G^+} (1-e^{i \rho \cdot \a})(1-e^{-i \rho \cdot \a}) = \frac{1}{|W_G|} \exp \left[ \sum_{n\ge1} \frac{1}{n} \left( - \chi_{\textbf{adj}} (n \alpha) + \textrm{rk}(G) \right)\right] \ ,
\end{align}
where $\textrm{rk}(G)$ is the rank of the gauge group and $|W_G|$ is the order of the Weyl group of $G$. We have the Cartan piece that is non-vanishing. There is no divergence coming from this term since the fermionic part of the Cartan contribution gives $-1+e^{n\Delta}$ to cancel the unity.
In the Cardy limit, this factor does contribute to the index, but it simply gives an overall volume factor of the form $(\frac{1}{\omega_1 \omega_2})^{\textrm{rk}(G)}$. 
Since this only makes a subleading logarithmic correction $ \textrm{log} (\omega_1 \omega_2)$ to our Cardy formula, we ignore this factor.
Therefore, the vector multiplet index (combined with the Haar measure) in the Cardy limit becomes
\begin{align}
    \label{eq:free-vec}
    - \frac{1}{\omega_1\omega_2} \sum_{\rho \in \Delta_G}\sum_{n\ge 1}\frac{(-e^{+\Delta + i\rho \cdot \alpha})^n - (-e^{-\Delta - i\rho \cdot \alpha})^n }{n^3} =  - \sum_{s = \pm} \frac{s}{\omega_1\omega_2} \sum_{\rho \in \Delta_G}\text{Li}_3(-e^{s(\Delta + i\rho \cdot \alpha)}).
\end{align}

For a chiral multiplet $\mathcal{X}$ in the representation $\mathbf{R}_{\mathcal{X}}$ of $G$, 
\begin{align}
    \CZ_{\mathcal{X}}\Big|_{\eqref{eq:constraint}} = \exp \left[\sum_n \frac{(-1)^{n-1}}{n}\sum_{w \in \mathbf{R}_{\mathcal{X}}} \left(\frac{t^{n(-r_\mathcal{X}+1)}e^{i n w ({\alpha})} - t^{n(r_\mathcal{X}-1)}e^{-i n w ({\alpha})} }{2\sinh(n\omega_1/2)\,2\sinh(n\omega_2/2)}\right)  \right].
\end{align}
Its exponent simplifies in the Cardy limit to the following expression:\footnote{
Had we studied the superconformal index of \cite{Kinney:2005ej, Romelsberger:2005eg} (with $(-1)^F$ insertion) instead of \eqref{eq:index} (with $e^{\pi i R}$ insertion), the formula \eqref{eq:free-chi} would have become
    \begin{align}
        \sum_{s=\pm }\frac{1}{\omega_1\omega_2} \sum_{\rho \in \mathbf{R}}\sum_{\lambda \in \mathbf{F}} \bigg[&s\, \text{Li}_3\,(e^{ i  s\rho ({\alpha}) -s \lambda ({m})}) + \omega_+(1-r_\mathcal{X})\text{Li}_2\,(e^{i  s\rho ({\alpha}) -s \lambda ({m})})
        \bigg] + \mathcal{O}(\omega^0)  \nonumber.
    \end{align}
    Ignoring the issue of holonomy saddles, all Li${}_3$ pairs become zero. Inserting Li${}_2(1) = \frac{\pi^2}{6}$ reproduces the asymptotic free energy of \cite{DiPietro:2014bca}, proportional to $\text{Tr}(R) \propto (a-c)$. See also \cite{Ardehali:2015bla}.}
\begin{align}
    \label{eq:free-chi}
    \sum_{s=\pm }\frac{s}{\omega_1\omega_2} \sum_{w \in \mathbf{R}_{\mathcal{X}}} \text{Li}_3(-e^{s(1-r_{\mathcal{X}})\Delta  +i  s w ({\alpha}) }) \ . 
\end{align}
Inserting \eqref{eq:free-vec} and \eqref{eq:free-chi} back to $\eqref{eq:sci-integral}$, the asymptotic expression of $\mathcal{I}$ becomes
\begin{align}
    \label{eq:sci-integral-cardy}
    \int [d\alpha] \exp{\left[\sum_{s = \pm} s \left(-\sum_{\rho \in \Delta_G} \frac{ \text{Li}_3(-e^{s(\Delta + i \rho \cdot \alpha)})}{\omega_1\omega_2} + \sum_\mathcal{\chi} \sum_{w \in \mathbf{R}_{\mathcal{X}}} \frac{ \text{Li}_3(-e^{s(1-r_{\mathcal{X}})\Delta  +i  s w ({\alpha}) })}{\w_1 \w_2} \right) \right]} , 
\end{align}
where $[d\a] =\prod_i d\a_i$. 

The holonomy integral \eqref{eq:sci-integral-cardy} can be performed by applying the saddle point approximation. The most dominant saddle point is located at the global minimum of the following expression:
\begin{align}
    \label{eq:exponent-cardy}
\sum_{s = \pm}  \frac{s}{\omega_1\omega_2}\left(\sum_{\rho \in \Delta_G}\text{Li}_3(-e^{s(\Delta + i \rho \cdot \alpha)}) - \sum_\mathcal{\chi} \sum_{w \in \mathbf{R}_{\mathcal{X}}} \text{Li}_3(-e^{s(1-r_{\mathcal{X}})\Delta  +i  s w ({\alpha})})\right) \equiv \frac{\mathcal{F}}{\omega_1\omega_2}
\end{align}
Now we search for the saddle point of the function $\CF$ to approximate the index integral in the Cardy limit. In the Cardy limit, the $U(1)_R$ chemical potential $\Delta$ should be $\Delta \sim -i \pi$ since $|w_+| \ll 1$. In addition, we assume that $\text{Im}\left({\omega_1\omega_2}\right) > 0$. The consistency of the assumption will be tested later in Section~\ref{sec:entropy}. 

We conjecture that the saddle point is at the origin in the gauge holonomies $\a_1 = \cdots = \a_{|G|} = 0$. One intuition behind this is as follows. When any of the holonomy variables get a non-zero value, gauge symmetry is (partially) broken so that we acquire some massive degrees of freedom. But we expect the high-temperature behavior for an asymptotically free gauge theory is in the maximally deconfining phase, rather than partially confining or Higgssed phase. Also, in the work of \cite{Aharony:2003sx}, such a maximally deconfining saddle was naturally assumed at high temperature, based on the intuitions from the solvable Gross-Witten-Wadia model \cite{Gross:1980he, Wadia:1980cp}.
This is rather different from the previous analysis \cite{DiPietro:2014bca, Ardehali:2015bla, Ardehali:2015hya, DiPietro:2016ond, Hwang:2018riu}, where the index does not capture the fully deconfining phase of the gauge theory due to the heavy bose-fermi cancellations. There one has to be careful about the non-trivial holonomy saddles \cite{Hwang:2017nop}, otherwise one gets an incorrect answer. In our case, we will see that the index captures enough degrees of freedom (eg. $\CO(e^{N^2})$ for the $SU(N)$ theory) to see the deconfining phase. Therefore it is natural to expect the `maximally deconfining' configuration dominates. In Section \ref{sec:examples}, we plot $(\text{Re}\mathcal{F},\,\text{Im}\mathcal{F})|_{\Delta \sim - i \pi}$ as a function of holonomies for various theories to support our claim. 

Assuming that the dominant saddle is indeed given at the origin, the asymptotic expression for the $\log{\mathcal{I}}$ can be written as 
\begin{align}
    \label{eq:free-cardy}
 \log(\CI) = \sum_{s = \pm}  \frac{s}{\omega_1\omega_2}\left(- |G|\,\text{Li}_3(-e^{s \Delta}) +\sum_{\mathcal{X}} |\mathbf{R}_{\mathcal{X}}| \, \text{Li}_3(-e^{s(1-r_{\mathcal{X}})\Delta })\right) .
\end{align}
One can further simplify this expression by applying the following Li${}_3$ identity,
\begin{align} \label{eq:Li3id}
    \text{Li}_3 (-e^{x}) - \text{Li}_3 (-e^{-x}) =  -\frac{(x-2\pi i p)^3}{6} -\frac{\pi^2 (x-2\pi i p)}{6},
\end{align}
which holds for $(2p-1)\pi < \text{Im}(x) < (2p+1)\pi$. The lower bound can be saturated when $\text{Re}(x) = 0$. Once we assume $0< r_\mathcal{X} < 2$ for any chiral multiplet\footnote{Many literatures on the subject (implicitly) assumes $0 < r_\CX < 2$ for the elementary fields \cite{Assel:2014paa, DiPietro:2014bca, Ardehali:2015bla}. This is required to put supersymmetric gauge theories on a three-sphere since a chiral multiplet has the conformal mass of the form $r_\CX (2-r_\CX)$. However, this condition is not always satisfied. For example, the gauge theories studied in \cite{Kutasov:1995np, Gadde:2015xta, Agarwal:2018ejn} have charged matter fields with $R$-charges less or equal to 0.  Nevertheless, the superconformal index for the fixed point is clearly well-defined beyond $0<r_\CX <2$. The superconformal index for the case with $r_\CX \le 0$ or $r_\CX \ge 2$ has been considered in \cite{Agarwal:2014rua, Agarwal:2015vla, Gadde:2015xta, Agarwal:2018ejn} for example. For these cases, we can turn on the chemical potentials for the (possibly anomalous) flavor symmetry to push the argument inside trilogarithm to be in the `canonical chamber' $|\textrm{Im}(x)|< \pi$. Then we turn off the flavor chemical potential to recover the index. Whenever this procedure can be done, our expression \eqref{eq:logIcanonical} gives the correct asymptotic limit of the index. For example, Kutasov-Schwimmer duality \cite{Kutasov:1995ve, Kutasov:1995np} sometimes maps a gauge theory with $R$-charge within 0 and 2 to a dual description with $r_\CX<0$ fields. Since the superconformal index for the two dual descriptions has to be identical, we claim our formula still holds for the case $R$ not within 0 and 2. See section 6 of \cite{Agarwal:2014rua} for a discussion on the integration contour issue for the index.}, $\Delta$ and $(1 - r_\CX ) \Delta$ appearing in \eqref{eq:free-cardy} are in the canonical chamber with their imaginary parts between $- \pi$ and $\pi$. 
Then \eqref{eq:free-cardy} becomes
\begin{align} \label{eq:logIcanonical}
    \textrm{log}(\CI) =  \frac{1}{\omega_1\omega_2} \bigg( \frac{|G|}{6}(\Delta^3 + \pi^2 \Delta) 
    - \sum_{\mathcal{X}} \frac{|\mathbf{R}_{\mathcal{X}}|}{6} \big((1-r_\mathcal{X})^3\Delta^3 + \pi^2 (1-r_\mathcal{X})\Delta\big)\bigg).
\end{align}
Using the following relation for a generic $\mathcal{N}=1$ gauge theory
\begin{align}
     \text{Tr}\,R^3 &= |G| +  \sum_{\mathcal{X}} (r_\mathcal{X}-1)^3 |\mathbf{R}_\mathcal{X}| = \frac{16}{9}(5a-3c) \ , \\
     \text{Tr}\,R &= |G| +  \sum_{\mathcal{X}} (r_\mathcal{X}-1) |\mathbf{R}_\mathcal{X}| = 16(a-c) \ , 
\end{align}
we can express the asymptotic free energy in terms of $c$ and $a$ as
 \begin{align}
    \label{eq:asymCardyF}
    \textrm{log}(\CI) =  \text{Tr}\,R^3 \,\frac{\Delta^3}{6\omega_1\omega_2} + \text{Tr}\,R \,\frac{\pi^2 \Delta}{6\omega_1\omega_2}  
     =  \frac{8(5a-3c)}{27\omega_1 \omega_2}\Delta^3 + \frac{8\pi^2(a-c)}{3\omega_1\omega_2}\Delta .
\end{align}
This is the key formula, which is reminiscent of the `entropy function' in $\CN=4$ SYM theory \cite{Hosseini:2017mds}. Indeed, it can be reduced to the $\CN=4$ formula upon taking $a=c$ and $\Delta_1=\Delta_2=\Delta_3$. Upon replacing $\Delta = \w_+ - i \pi$, and writing the leading powers in $\w$, we obtain
\begin{align}
    \textrm{Re}(\textrm{log}(\CI)) = (3c-2a)\frac{16 \pi^2 \textrm{Im}(\w_1 \w_2)}{27 |\w_1|^2 |\w_2|^2} + \CO(\w^{-1}) \ . 
\end{align}
We see that this indeed reproduces the $N^2$ growth for the `high-temperature' limit of the $SU(N)$ $\CN=4$ SYM theory with $a=c \sim N^2$. As long as we choose the imaginary part of $\w_1 \w_2$ to be positive, we find the exponential growth of the states in the `high-temperature' limit controlled by the combination of central charges $3c-2a$. This is enough to account for the $\CO(N^2)$ growth of $\CN=4$ SYM theory which has $a=c\sim N^2$ \cite{Choi:2018hmj}. 

So far we did not specify the precise value of $\w_{1, 2}$. In section \ref{sec:entropy} we will perform the Legendre transformation to entropy and then extremize it with respect to chemical potentials $\w_i$. We find that at the extremum, $\textrm{Im}(\w_1 \w_2)>0$. Also, if $\w_1 = \w_2 = \w$, we get $\textrm{Re}(\w) \simeq \sqrt{3} \textrm{Im}(\w) $ so that the $\w$ has the phase near $\pi/6$. 

\paragraph{Index with flavor chemical potentials}
Let us slightly generalize our Cardy formula by including the chemical potentials $x = (x^1, x^2, \ldots, x^n)$ for the flavor symmetry. Let us assume, for simplicity, that the flavor symmetries are abelian $U(1)^n$ and denote the flavor generators as $F_I$ with $I=1, 2, \ldots, n$. For a chiral multiplet of representation $\mathbf{R}$ under the gauge group $G$ and the flavor charge $Q=(Q_1, \ldots, Q_n)$, the index becomes
\begin{align}
& \CZ_{\CX}  \simeq \left[ \frac{1}{\w_1 \w_2} \sum_{n \ge 1} \frac{(-1)^{n-1}}{n^3} \sum_{w \in \mathbf{G}} \left(  t^{n(1-r_\CX )} e^{n(i w (\a) + i Q \cdot x))} - t^{n(r_\CX-1)} e^{-n(i w(\a) + i Q\cdot x)} \right) \right] \nn \\
 &~= \exp \left[ -\frac{1}{\w_1 \w_2} \sum_{w \in \mathbf{R}} \left\{ \textrm{Li}_3 \left(-e^{(1-r_\CX) \Delta + i w (\a) + i Q\cdot x}\right) - \textrm{Li}_3 \left(-e^{-(1-r_\CX) \Delta - i w (\a) - i Q\cdot x } \right) \right\} \right] \\
 &~= \exp \left[ \frac{1}{6 \w_1 \w_2} \sum_{w \in \mathbf{R}} \left\{ \Big( (1-r_\CX)\Delta + i w (\a) + i Q\cdot x \Big)^3 + {\pi^2}\Big( (1-r_\CX)\Delta + i w (\a) + i Q\cdot x \Big) \right\} \right] . \nn
\end{align}
Assuming the dominant saddle point is located at the origin $\a_1 = \a_2= \cdots = 0$, the index for a gauge theory in the Cardy limit can be written as
\begin{align} 
 \CI &\sim \exp \left[  \sum_{\a \in \Delta} \frac{\Delta^3 + \pi^2 \Delta }{6\w_1 \w_2} + 
 \sum_{\CX} \left( \frac{ |\mathbf{R}_\CX |}{6\w_1 \w_2 } (i Q_\CX \cdot x + (r_\CX-1)\Delta)^3 + \frac{\pi^2  |\mathbf{R}_\CX |}{6 \w_1 \w_2} (i Q_\CX \cdot x + (r_\CX-1)\Delta) \right)  \right]  \nn \\
 &= \exp \left[ \frac{\Delta^3}{6w_1 w_2}\left( |G| + \sum_\CX (r_\CX-1)^3 |\mathbf{R}_\CX| \right) + \frac{\pi^2 \Delta }{6\w_1 \w_2} \left( |G| + \sum_\CX (r_\CX-1) |\mathbf{R}_\CX| | \right) \right]  \\
 & \qquad \times \exp \left[ \frac{i \Delta^2 }{2 \w_1 \w_2} \sum_{\CX} |\mathbf{R}_\CX|  Q_\CX \cdot x (r_\CX -1)^2 
 - \frac{\Delta}{2\w_1 \w_2} \sum_{\CX}  |\mathbf{R}_\CX|  (Q_\CX \cdot x)^2 (r_\CX - 1)  \right] \nn \\
 & \qquad \times \exp \left[  \frac{i \pi^2}{6 \w_1\w_2} \sum_{\CX} |\mathbf{R}_\CX| Q_\CX \cdot x - \frac{i}{6\w_1 \w_2} \sum_{\CX} |\mathbf{R}_\CX| (Q_\CX \cdot x)^3 \right] \nn
\end{align}
Now we can use the trace anomalies to simplify the above formula:
\begin{align}
\begin{split}
 k_{RRR} \equiv \Tr R^3 &= |G| + \sum_{\CX} (r_\CX-1)^3 |\mathbf{R}_\CX| \ , \\ 
 k_R \equiv  \Tr R &= |G| + \sum_{\CX} (r_\CX-1) |\mathbf{R}_\CX| \ ,\\
 k_{RRI} \equiv \Tr R^2 F_I &= \sum_{\CX} (r_\CX - 1)^2 |\mathbf{R}_\CX| Q_{\CX, I} \ , \\
 k_{RIJ}  \equiv  \Tr R F_I F_J &= \sum_{\CX} |\mathbf{R}_\CX| (r_\CX-1) Q_{\CX, I} Q_{\CX, J} \ , \\
 k_{IJK} \equiv \Tr F_I F_J F_K &= \sum_{\CX} |\mathbf{R}_\CX| Q_{\CX, I} Q_{\CX, J} Q_{\CX, K}   \ , \\
 k_{I} \equiv \Tr F_I &= \sum_{\CX}  |\mathbf{R}_\CX| Q_{\CX, I} \ .
\end{split}
\end{align}
Here $Q_{\CX, I}$ denotes the $U(1)_I$ charge of the chiral multiplet $\CX$. For the superconformal field theory, we have the relations $k_I = 9k_{RRI}$ and $k_{RIJ} < 0$ \cite{Osborn:1998qu, Intriligator:2003jj}. Then the index in the Cardy limit can be written as
\begin{align}
\log \CI \sim \frac{k_{RRR}  \Delta^3 + 3i k_{RRI} \Delta^2 x^I - 3k_{RIJ} \Delta x^I x^J  - i k_{IJK} x^I x^J x^K }{6\w_1 \w_2} + \frac{\pi^2 (k_R \Delta  + i k_I x^I) }{6\w_1 \w_2} . 
\end{align}
When the flavor symmetry is baryonic, $k_I = k_{RRI} = 0$ so that the sum involving $k_I$ or $k_{RRI}$ only runs over non-baryonic flavor symmetries. 

\section{Background field method on $S^3$} \label{sec:eff-action}

So far, we have relied upon the explicit expression for the supersymmetric index of a Lagrangian theory. In this section, we describe how to obtain the Cardy free energy \eqref{eq:asymCardyF} without referring to the Lagrangian description of a 4d $\mathcal{N}=1$ SCFT as was done in \cite{DiPietro:2014bca}. Let us consider the effective action of the background fields coupled to the SCFT on $S^3 \times S^1$. The chemical potentials $\beta$, $\omega_1$, $\omega_2$, $\Delta$ introduced in \eqref{eq:Zdef} should appear as the 4d metric and background gauge fields as follows:
\begin{align}
    ds^2 = r^2\left[d\theta^2 + \sum_{i=1}^2 n_i^2 \left(d\phi_i - \frac{i\omega_i}{\beta}\,d\tau\right)^2
    \right] + d\tau^2, \quad    A = \frac{i\Delta}{\beta}d\tau.
\end{align}
Since we are interested in the small circle limit $\beta/r \ll 1$, it is convenient to arrange the above 4d background in terms of the 3d background fields, i.e., $ds^2 = ds^2_3 + e^{-2\Phi}(d\tau + a)^2$ and  $A = A_4(d\tau+a) + \mathcal{A}$.
The metric $ds^2_3$, graviphoton $a$, dilaton $e^{-2\Phi}$, $U(1)_R$ connection $\mathcal{A}$, scalar $A_4$ are given as
\begin{gather}
    \label{eq:background-field-value}
    \begin{split}
    ds^2_3 &= r^2\left[d\theta^2 + \sum_{n=1}^2 n_i^2d\phi_i^2 +\frac{r^2(\sum_i \omega_i n_i^2 d\phi_i)^2}{\beta^2(1-r^2\sum_i \frac{n_i^2\omega_i^2}{\beta^2})}\right], \quad a = -i\frac{r^2\sum_i \omega_i n_i^2 d\phi_i}{\beta(1-r^2\sum_i \frac{n_i^2\omega_i^2}{\beta^2})},\\[0.1in]
    & \quad\quad\quad e^{-2\Phi} = 1-r^2\sum_{i}\frac{n_i^2\omega_i^2}{\beta^2}, \qquad
    A_4 = \frac{i\Delta}{\beta}, \qquad \mathcal{A} = -A_4 a.
    \end{split}
\end{gather}

Firstly, let us consider Lagrangian theories. We consider the path integral expression of the index \eqref{eq:Zdef} in this background. Each 4d field can be separated into the 3d zero mode and the Kaluza-Klein  tower. Since the 3d QFT is well-defined in UV, the only possible divergence in $|\omega| \ll 1$ should be the IR divergence, contributing to the free energy
at the subleading order $\mathcal{O}(\log{\omega})$ \cite{Choi:2018hmj}. To find the leading free energy in the Cardy limit, therefore, it is sufficient to integrate out all the Kaluza-Klein modes of 4d dynamical fields. All 4d fermions are anti-periodic before imposing the constraint \eqref{eq:constraint}. In a generic 4d background, the masses of the KK fermions are shifted by $A_4 = i\Delta/\beta$, such that $m_n = n-\frac{\beta}{2\pi}(1-R)A_4$ for $n \in \mathbb{Z}+\frac{1}{2}$, where $(1-R)$ is the $U(1)_R$ charge of the fermion. Integrating out the KK fermion generates the effective Chern-Simons action. Unless there is a massless fermion in the KK tower, i.e., $-\frac{1}{2}< \frac{\beta }{2\pi}(1-R)A_4  < \frac{1}{2}$, the CS action is \cite{DiPietro:2014bca}
\begin{align}
    \label{eq:action-CS}
     -\frac{i\beta\, (1-R)^3}{2(2\pi)^2} \int  \left(A_4 \mathcal{A} \wedge d\mathcal{A} +  A_4^2 \mathcal{A}\wedge da + \frac{1}{3}A_4^3 a\wedge da\right) -\frac{i\,(1-R)}{2\beta}\int \left(\frac{1}{12}\mathcal{A} \wedge da\right) . 
\end{align}
Given the BPS index condition \eqref{eq:constraint}, the Cardy limit $|\omega| \ll 1$ inevitably implies $\Delta \sim -i \pi$. Here we assume the $U(1)_R$ charge of every chiral multiplet $\mathcal{X}$ is in the range of $0<R_{\mathcal{X}}<2$ to avoid the appearance of massless fermions.\footnote{This assumption $0<R_{\mathcal{X}}<2$ is not always true as we discussed in the footnote near \eqref{eq:logIcanonical}. Given our expression does not depend on a specific choice of UV gauge theory or dual descriptions, we conjecture that \eqref{eq:action-CS} is correct for arbitrary superconformal theories. 
} Summing over all the fermions in the theory, we find \cite{DiPietro:2014bca}
\begin{align}
    \label{eq:action-CS}
    \begin{split}
    S_{\rm CS} &= S_{\rm cubic} + S_{\rm mixed} , \\
    S_{\rm cubic} &=  -\frac{i\beta\, \text{Tr}(R^3)}{2(2\pi)^2} \int  \left(A_4 \mathcal{A} \wedge d\mathcal{A} +  A_4^2 \mathcal{A}\wedge da + \frac{1}{3}A_4^3 a\wedge da\right) , \\
     S_{\rm mixed} &=  -\frac{i\,\text{Tr}(R)}{2\beta}\int \left(\frac{1}{12}\mathcal{A} \wedge da\right).
     \end{split}
\end{align}
We can write the trace anomalies in terms of central charges as $\Tr R^3 = \frac{16}{9}(5a-3c)$ and $\Tr R = 16(c-a)$. 

One can also understand the CS action \eqref{eq:action-CS} from an abstract anomaly matching \cite{Banerjee:2012iz,Jensen:2013rga,DiPietro:2014bca,Golkar:2015oxw} without referring to Lagrangian. The first term  $S_{\rm cubic}$ is not invariant under the background $U(1)_R$ gauge transformation. The presence of this term is required to match the 4d $U(1)_R^3$ covariant anomaly \cite{Banerjee:2012iz,Jensen:2013rga,DiPietro:2014bca}. 
Matching the mixed $U(1)_R$ gauge-gravity anomaly also requires the inclusion of another Chern-Simons term. However, the corresponding term
must contain 3 derivatives, thus being proportional to $\beta^3$. It is suppressed in the limit $\beta/r \ll |\omega_{1,2}| \ll 1$. 
The gauge-invariant Chern-Simons term, $S_{\rm mixed}$, is required for anomaly matching under the large gauge transformations and the large diffeomorphisms \cite{Jensen:2013rga,Golkar:2015oxw}. Other gauge-invariant Chern-Simons terms, such as $\int a \wedge da$ or $\int \mathcal{A} \wedge d\mathcal{A}$, are not allowed due to the $\mathcal{C}\mathcal{P}\mathcal{T}$ invariance of 4d QFT \cite{Banerjee:2012iz}. Therefore, the expression \eqref{eq:action-CS} can be derived without referring to a Lagrangian description. 

There are infinitely many possible terms in constructing the background fields' effective action, apart from the Chern-Simons terms. It was found in \cite{Choi:2018hmj} that all terms which involve the volume integral $\int d^3 x\sqrt{g}$ are suppressed in the BPS and the Cardy limit, i.e., $\beta/r \ll \omega_{1,2} \ll 1$. 
Likewise, those terms which involve the totally antisymmetric tensor $\epsilon^{\mu\nu\sigma}$ in the Lagrangian density are suppressed in the Cardy limit if they do not belong to \eqref{eq:action-CS} \cite{Choi:2018hmj}. In summary, the only non-vanishing possibilities are the gauge invariant and non-invariant Chern-Simons terms \eqref{eq:action-CS}.

Finally, we plug in the actual value \eqref{eq:background-field-value} of background fields into the action \eqref{eq:action-CS}. The evaluated action $S_{\rm cubic}$ and $S_{\rm mixed}$ are in agreement with \eqref{eq:asymCardyF}, i.e.,
\begin{align}
\begin{split}
    S_{\rm cubic} &= -\text{Tr}(R^3) \,\frac{\Delta^3}{6\omega_1\omega_2} = - \frac{8 (5a-3c) \Delta^3}{27 \w_1 \w_2},  \\
    S_{\rm mixed} &= -\text{Tr}(R) \,\frac{\pi^2 \Delta}{6\omega_1\omega_2} = -  \frac{16 \pi^2 (a-c) \Delta}{\w_1 \w_2},
\end{split}
\end{align}
since 
\begin{align}
    \int a\wedge da =   \frac{(2\pi)^2 r^4 \omega_1\omega_2}{\beta^2(1-\frac{r^2 \omega_1^2}{\beta^2})(1-\frac{r^2 \omega_2^2}{\beta^2})} \simeq \frac{(2\pi)^2\beta^2}{\omega_1\omega_2} + \mathcal{O}\left(\frac{\beta^4}{r^2\omega^4}\right).
\end{align}
This result perfectly agrees with our previous computation using the free field theory analysis of the index.

\section{Saddle point analysis} \label{sec:examples}

We assumed in Section~\ref{subsec:asymptotic-index}~and~\ref{sec:eff-action} that the most dominant saddle point of the holonomy integral \eqref{eq:sci-integral-cardy} is at the origin $\alpha_1 = \alpha_2 = \cdots = \alpha_{|G|}=0$. 
In this section, we provide supporting evidence for the prior assumption. We numerically search for the most dominant saddle point for a large set of theories: SQCDs with different gauge groups and flavors, $\mathcal{N}=4$ SYM with different gauge groups, ISS model \cite{Intriligator:1994rx}, BCI model \cite{Brodie:1998vv}, Pouliot theory \cite{Pouliot:1995zc} and $SU(2)^3$ gauge theory coupled via trifundmental chiral multiplets. We also consider `non-Lagrangian' Argyres-Douglas theories of $(A_1, A_N)$ type \cite{Argyres:1995xn,Argyres:1995jj, Eguchi:1996vu} using the $\mathcal{N}=1$ Lagrangian description \cite{Maruyoshi:2016tqk, Maruyoshi:2016aim}. This set of examples includes the theories with $c<a$ which were shown to have non-trivial holonomy saddle points \cite{Ardehali:2015bla} when the chemical potentials are real-valued.

We have also investigated the possibility of a saddle point away from the real line, having a complex value. 
Before taking the Cardy limit, there are infinitely many poles inside the integration contour. As we take the Cardy limit, these poles collide to form a branch cut. Therefore we cannot move the contour away from the origin without changing the value of the integral. At least for the case of $SU(2)$ gauge theory, we explicitly verified that there is no other saddle point besides the one at the origin. Throughout the rest of this section, we focus on the real values of $\a$ assuming that there is no other saddle point, or any possible complex saddle is sub-dominant.

We make the following assumption throughout this section:
\begin{align}
    \label{eq:assumption}
    \text{Im}\left(\omega_1\omega_2\right) > 0
\end{align}
The consistency of this assumption is tested in Section~\ref{sec:entropy}, by finding the actual solution of the extremization equation \eqref{eq:extreme-sol} at $\omega_1 = \omega_2$. The asymptotical value of the $U(1)_R$ chemical potential is  $\Delta \sim -i \pi$ due to the constraint $\Delta = -i\pi  + \omega_+$. So we evaluate the numerical value of $(\text{Re}\mathcal{F},\,\text{Im}\mathcal{F})|_{\Delta \sim -i \pi}$ on the gauge holonomy space. Mathematica's \texttt{NMaximize} and \texttt{NMinimize} functions can be used to show that
\begin{enumerate}[label=(\roman{*}), ref=(\roman{*})]
    \setlength\itemsep{-0.2em}
    \item \label{itm:cond1} $\text{Re}\mathcal{F}|_{\Delta = -i \pi} = 0$. More generally, $\text{Re}\mathcal{F}$ is nearly zero at $\Delta \sim -i (\pi-\varepsilon) \pm \varepsilon$ with $\varepsilon \ll 1$.
    \item \label{itm:cond2} $\text{Im}\mathcal{F}$ has a global minimum at $\alpha_1 = \cdots = \alpha_{|G|}=0$ for $\Delta \sim -i (\pi-\varepsilon) \pm \varepsilon$ with $\varepsilon \ll 1$.\footnote{Generally, it is a challenging task to find the global minimum/maximum in a multi-dimensional space. Mathematica's \texttt{NMinimize} or \texttt{NMaximize} sometimes fails to identify the global extremum and only finds local extrema. Whenever \texttt{NMinimize} identifies a minimum point $\vec{p} \neq 0$, we checked $\text{Im}\mathcal{F}|_{\vec{\alpha} = 0} < \text{Im}\mathcal{F}|_{\vec{\alpha} = \vec{p}}$ at least.}
\end{enumerate}
The above two conditions, combined with \eqref{eq:assumption}, should be sufficient to conclude that the most dominant saddle point of the integral \eqref{eq:sci-integral-cardy} is located at the origin.

\begin{figure}[t!]
    \centering
    \begin{subfigure}[b]{0.23\textwidth}\centering
        \includegraphics[width=\textwidth]{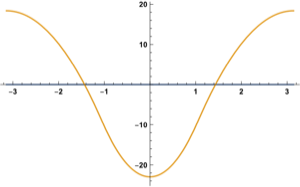}
        \caption{$SU(2)$ $N_f=3$}
        \label{fig:su2nf3}
    \end{subfigure}\hspace{0.2cm}
    \begin{subfigure}[b]{0.23\textwidth}\centering
        \includegraphics[width=\textwidth]{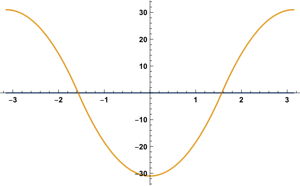}
        \caption{$SU(2)$ $N_f=4$}
        \label{fig:su2nf4}
    \end{subfigure}\hspace{0.2cm}
    \begin{subfigure}[b]{0.23\textwidth}\centering
        \includegraphics[width=\textwidth]{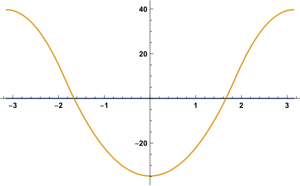}
        \caption{$SU(2)$ $N_f=5$}
        \label{fig:su2nf5}
    \end{subfigure}\hspace{0.2cm}
    \begin{subfigure}[b]{0.23\textwidth}\centering
        \includegraphics[width=\textwidth]{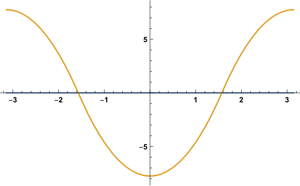}
        \caption{$SO(3)$ $N_f=2$}
        \label{fig:so3nf2}
    \end{subfigure}\\
    \caption{$(\text{Re}\mathcal{F},\,\text{Im}\mathcal{F})|_{\Delta = -i \pi}$ of rank-1 SQCDs. The blue/orange line is $\text{Re}\mathcal{F}$/$\text{Im}\mathcal{F}$.}
    \label{fig:rank1}
\end{figure}

\begin{figure}[!htbp]
    \centering
    \begin{subfigure}[b]{0.2\textwidth}\centering
        \includegraphics[width=\textwidth]{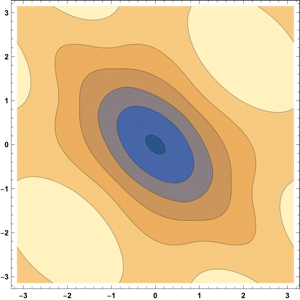}
        \caption{$SU(3)$ $N_f=4$}
        \label{fig:SU3NF4}
    \end{subfigure}\hspace{0.8cm}
    \begin{subfigure}[b]{0.2\textwidth}\centering
        \includegraphics[width=\textwidth]{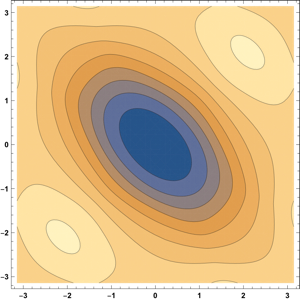}
        \caption{$SU(3)$ $N_f=5$}
        \label{fig:SU3NF5}
    \end{subfigure}\hspace{0.8cm}
    \begin{subfigure}[b]{0.2\textwidth}\centering
        \includegraphics[width=\textwidth]{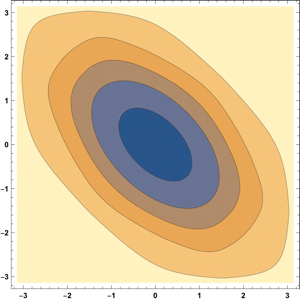}
        \caption{$SU(3)$ $N_f=6$}
        \label{fig:SU3NF6}
    \end{subfigure}\hspace{0.8cm}
    \begin{subfigure}[b]{0.2\textwidth}\centering
        \includegraphics[width=\textwidth]{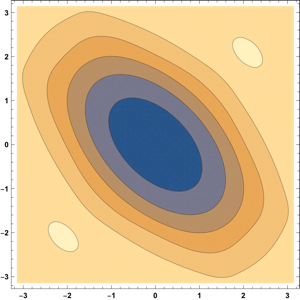}
        \caption{$SU(3)$ $N_f=7$}
        \label{fig:SU3NF7}
    \end{subfigure}\\\vspace{0.2cm}
    \begin{subfigure}[b]{0.2\textwidth}\centering
        \includegraphics[width=\textwidth]{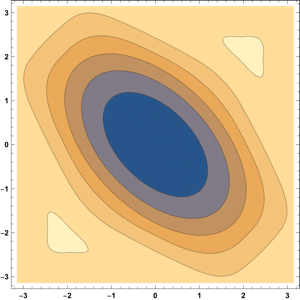}
        \caption{$SU(3)$ $N_f=8$}
        \label{fig:SU3NF8}
    \end{subfigure}\hspace{0.8cm}
    \begin{subfigure}[b]{0.2\textwidth}\centering
        \includegraphics[width=\textwidth]{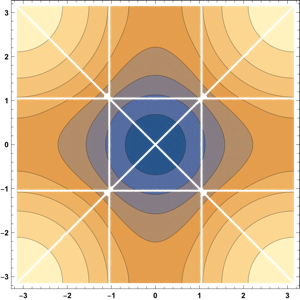}
        \caption{$SO(4)$ $N_f=3$}
        \label{fig:SO4NF3}
    \end{subfigure}\hspace{0.8cm}
    \begin{subfigure}[b]{0.2\textwidth}\centering
        \includegraphics[width=\textwidth]{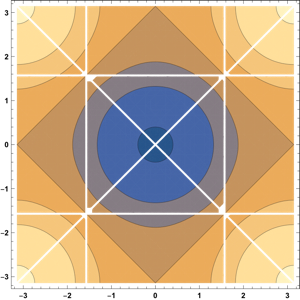}
        \caption{$SO(4)$ $N_f=4$}
        \label{fig:SO4NF4}
    \end{subfigure}\hspace{0.8cm}
    \begin{subfigure}[b]{0.2\textwidth}\centering
        \includegraphics[width=\textwidth]{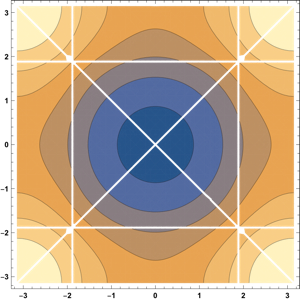}
        \caption{$SO(4)$ $N_f=5$}
        \label{fig:SO4NF5}
    \end{subfigure}\\\vspace{0.2cm}
    \begin{subfigure}[b]{0.2\textwidth}\centering
        \includegraphics[width=\textwidth]{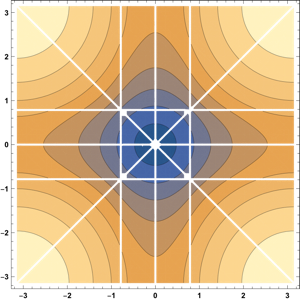}
        \caption{$SO(5)$ $N_f=4$}
        \label{fig:SO5NF4}
    \end{subfigure}\hspace{0.8cm}
    \begin{subfigure}[b]{0.2\textwidth}\centering
        \includegraphics[width=\textwidth]{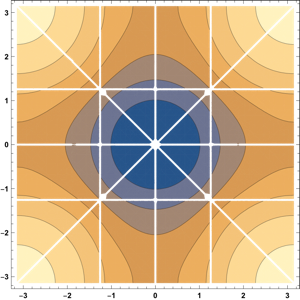}
        \caption{$SO(5)$ $N_f=5$}
        \label{fig:SO5NF5}
    \end{subfigure}\hspace{0.8cm}
    \begin{subfigure}[b]{0.2\textwidth}\centering
        \includegraphics[width=\textwidth]{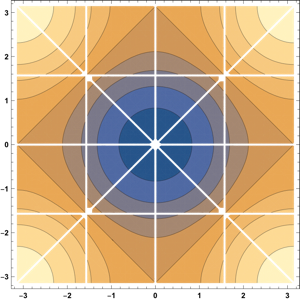}
        \caption{$SO(5)$ $N_f=6$}
        \label{fig:SO5NF6}
    \end{subfigure}\hspace{0.8cm}
    \begin{subfigure}[b]{0.2\textwidth}\centering
        \includegraphics[width=\textwidth]{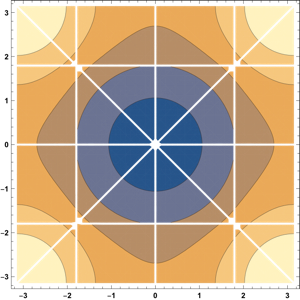}
        \caption{$SO(5)$ $N_f=7$}
        \label{fig:SO5NF7}
    \end{subfigure}\\\vspace{0.2cm}
    \begin{subfigure}[b]{0.2\textwidth}\centering
        \includegraphics[width=\textwidth]{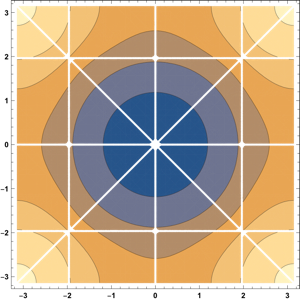}
        \caption{$SO(5)$ $N_f=8$}
        \label{fig:SO5NF8}
    \end{subfigure}\hspace{0.8cm}
    \begin{subfigure}[b]{0.2\textwidth}\centering
        \includegraphics[width=\textwidth]{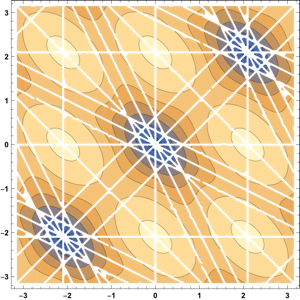}
        \caption{$G_2$ $N_f=5$}
        \label{fig:G2NF5}
    \end{subfigure}\hspace{0.8cm}
    \begin{subfigure}[b]{0.2\textwidth}\centering
        \includegraphics[width=\textwidth]{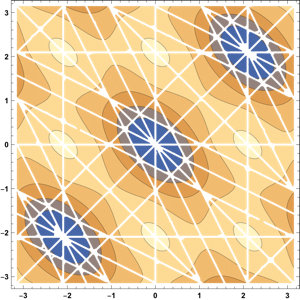}
        \caption{$G_2$ $N_f=6$}
        \label{fig:G2NF6}
    \end{subfigure}\hspace{0.8cm}
    \begin{subfigure}[b]{0.2\textwidth}\centering
        \includegraphics[width=\textwidth]{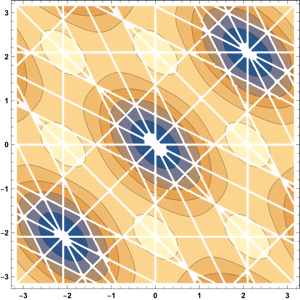}
        \caption{$G_2$ $N_f=7$}
        \label{fig:G2NF7}
    \end{subfigure}\\\vspace{0.2cm}
    \begin{subfigure}[b]{0.2\textwidth}\centering
        \includegraphics[width=\textwidth]{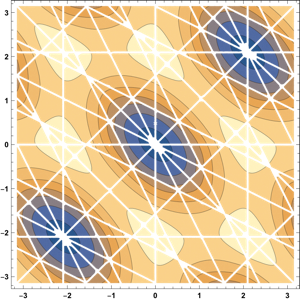}
        \caption{$G_2$ $N_f=8$}
        \label{fig:G2NF8}
    \end{subfigure}\hspace{0.8cm}
    \begin{subfigure}[b]{0.2\textwidth}\centering
        \includegraphics[width=\textwidth]{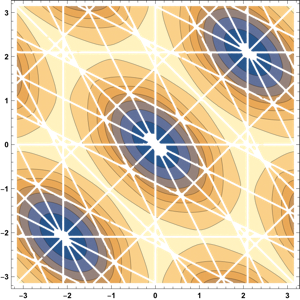}
        \caption{$G_2$ $N_f=9$}
        \label{fig:G2NF9}
    \end{subfigure}\hspace{0.8cm}
    \begin{subfigure}[b]{0.2\textwidth}\centering
        \includegraphics[width=\textwidth]{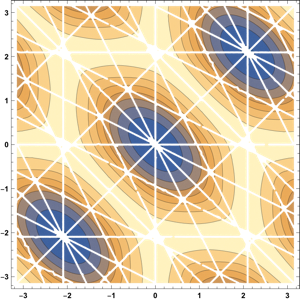}
        \caption{$G_2$ $N_f=10$}
        \label{fig:G2NF10}
    \end{subfigure}\hspace{0.8cm}
    \begin{subfigure}[b]{0.2\textwidth}\centering
        \includegraphics[width=\textwidth]{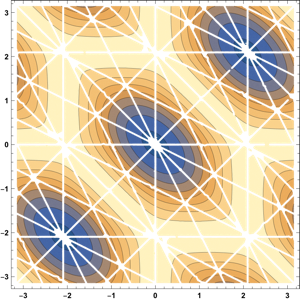}
        \caption{$G_2$ $N_f=11$}
        \label{fig:G2NF11}
    \end{subfigure}
    \caption{The contour plots of $\text{Im}\mathcal{F}|_{\Delta = -i \pi}$ for rank-2 SQCDs. The brighter/darker region has bigger/smaller value. The $Sp(2)$ plots are omitted since $\CF_{Sp(2)}^{N_f} - \CF_{SO(5)}^{N_f} = \text{(const)}$.
    The white lines are located at the cusps at which the $\text{Li}_3$ function jumps between the branches of \eqref{eq:Li3id}. The function is still smooth, and no additional saddle point exists on those lines. 
    }
    \label{fig:rank2}
\end{figure}

\paragraph{SQCD}
We first consider SQCDs with various gauge groups. There are $N_f$ fundamental and anti-fundamental chiral multiplets in $SU(N)$ and $E_6$ SQCDs. The $Sp(N)$ SQCD has an even number ($2N_f$) of fundamental chiral multiplets to avoid the Witten anomaly \cite{Witten:1982fp}.  The SQCDs with other gauge groups have $N_f$ fundamental chiral multiplets. The $R$-charge of the chiral multiplet for each gauge group is fixed to be
\begin{align}
    R_{SU(N)} &= 1-\frac{N}{N_f}, &  R_{SO(N)} &= 1-\frac{N-2}{N_f}, &  R_{Sp(N)} &= 1-\frac{N+1}{N_f}, & R_{E_6} &= 1-\frac{2}{N_f},\nn \\  
    R_{E_7} &= 1-\frac{3}{N_f},& R_{E_8} &= 1-\frac{1}{N_f}, &  R_{F_4} &= 1 - \frac{3}{N_f}, &  R_{G_2} &= 1-\frac{4}{N_f} . 
\end{align}
We have performed the analysis up to rank-10 gauge groups for the following range of $N_f$, whose IR fixed point corresponds to either an interacting SCFT or a free theory:
\begin{align}
    &N < N_f^{SU(N)} < 3N, \quad (N-2) < N_f^{SO(N)} < 3(N-2), \quad (N+1) < N_f^{Sp(N)} < 3(N+1) \nonumber\\
    &2 < N_f^{E_6} < 6,   \ \quad 
    3 < N_f^{E_7} < 9  , \ \quad 1 < N_f^{E_8} < 3  , \ \quad 3 < N_f^{F_4} < 9 ,\ \quad 4 < N_f^{G_2} < 12. 
\end{align}
Notice that they do not necessarily belong to the conformal window. 
By examining the real and imaginary value of $\mathcal{F}$, given as
\begingroup
\allowdisplaybreaks
\begin{align}
    \mathcal{F} = 
        \sum_{s = \pm}  s\Big(&
        \sum_{\rho \in \mathbf{adj}} \text{Li}_3(-e^{s\Delta + i s \rho (\vec{\alpha})} ) - N_f    \sum_{\lambda \in \mathbf{fnd}}
        \text{Li}_3(-e^{s(1-R)\Delta  +i  s \lambda(\vec\alpha)})\\
         & \qquad \qquad \qquad - N_f \sum_{\lambda \in \overline{\mathbf{fnd}}} \text{Li}_3(-e^{s(1-R)\Delta  +i  s \lambda(\vec\alpha)}) \Big)&& \text{ if }G = SU(N), E_6 \nn
        \\[0.0cm]
    \mathcal{F} = 
        \sum_{s = \pm}  s\Big(& \sum_{\rho \in \mathbf{adj}} \text{Li}_3(-e^{s\Delta + i s \rho (\vec{\alpha})} ) 
        - 2N_f \sum_{\lambda \in \mathbf{fnd}} \text{Li}_3(-e^{s(1-R)\Delta  +i  s \lambda(\vec\alpha)})   \Big) && \text{ if } G= Sp(N) \nn \\[0.0cm]
    \mathcal{F} = 
        \sum_{s = \pm}  s\Big(& \sum_{\rho \in \mathbf{adj}} \text{Li}_3(-e^{s\Delta + i s \rho (\vec{\alpha})} ) 
        - N_f \sum_{\lambda \in \mathbf{fnd}}\text{Li}_3(-e^{s(1-R)\Delta  +i  s \lambda(\vec\alpha)})   \Big) && \text{ otherwise}, \nn
\end{align}
\endgroup
we find that the conditions \ref{itm:cond1}~and~\ref{itm:cond2} are satisfied, thus $\alpha_1 = \cdots = \alpha_{|G|} = 0$ being the most dominant saddle point of \eqref{eq:sci-integral-cardy}. For rank-1 SQCD theories, $(\text{Re}\mathcal{F},\,\text{Im}\mathcal{F})|_{\Delta = -i \pi}$ are plotted in Figure~\ref{fig:rank1}. See also Figure~\ref{fig:rank2} for the contour plots of $\text{Im}(\mathcal{F})|_{\Delta = -i\pi}$ for rank-2 SQCDs.

\afterpage{\clearpage}

\begin{figure}[!t]
    \centering
    \begin{subfigure}[b]{0.45\textwidth}\centering
        \includegraphics[width=\textwidth]{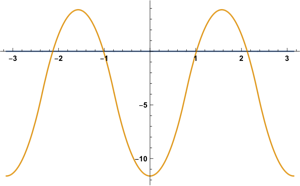}
        \caption{$SU(2)$ $N_a=2$}
        \label{fig:SU2NA2}
    \end{subfigure}\hspace{1cm}
    \begin{subfigure}[b]{0.45\textwidth}\centering
        \includegraphics[width=\textwidth]{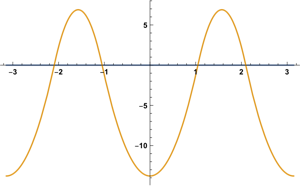}
        \caption{$SU(2)$ $N_a=3$}
        \label{fig:SU2NA3}
    \end{subfigure}\\
    \caption{$(\text{Re}\mathcal{F},\,\text{Im}\mathcal{F})|_{\Delta = -i \pi}$ of rank-1 SYM with $N_a=2,3$ adjoint chiral multiplets.}
    \label{fig:rank1msym}
\end{figure}

\begin{figure}[!htbp]
    \centering
    \begin{subfigure}[b]{0.23\textwidth}\centering
        \includegraphics[width=\textwidth]{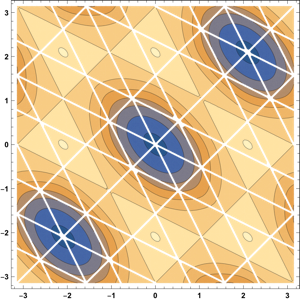}
        \caption{$SU(3)$ $N_a=2$}
        \label{fig:SU3NA2}
    \end{subfigure}\hspace{0.2cm}
    \begin{subfigure}[b]{0.23\textwidth}\centering
        \includegraphics[width=\textwidth]{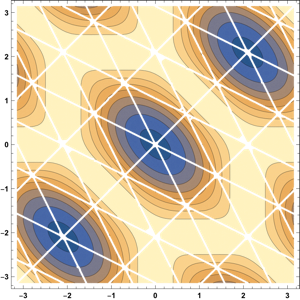}
        \caption{$SU(3)$ $N_a=3$}
        \label{fig:SU3NA3}
    \end{subfigure}\hspace{0.2cm}
    \begin{subfigure}[b]{0.23\textwidth}\centering
        \includegraphics[width=\textwidth]{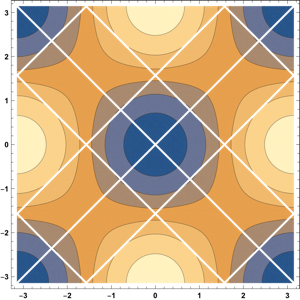}
        \caption{$SO(4)$ $N_a=2$}
        \label{fig:SO4NA2}
    \end{subfigure}\hspace{0.2cm}
    \begin{subfigure}[b]{0.23\textwidth}\centering
        \includegraphics[width=\textwidth]{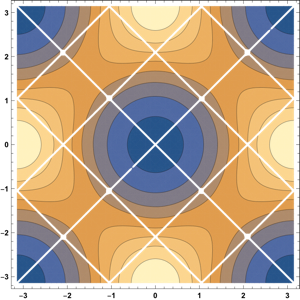}
        \caption{$SO(4)$ $N_a=3$}
        \label{fig:SO4NA3}
    \end{subfigure}\\\vspace{0.2cm}
    \begin{subfigure}[b]{0.23\textwidth}\centering
        \includegraphics[width=\textwidth]{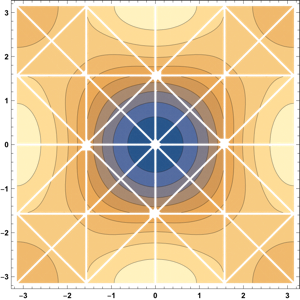}
        \caption{$SO(5)$ $N_a=2$}
        \label{fig:SO5NA2}
    \end{subfigure}\hspace{0.2cm}
    \begin{subfigure}[b]{0.23\textwidth}\centering
        \includegraphics[width=\textwidth]{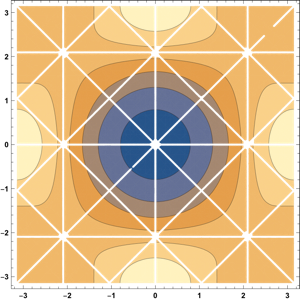}
        \caption{$SO(5)$ $N_a=3$}
        \label{fig:SO5NA3}
    \end{subfigure}\hspace{0.2cm}
    \begin{subfigure}[b]{0.23\textwidth}\centering
        \includegraphics[width=\textwidth]{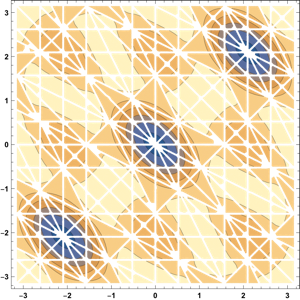}
        \caption{$G_2$ $N_a=2$}
        \label{fig:G2NA2}
    \end{subfigure}\hspace{0.2cm}
    \begin{subfigure}[b]{0.23\textwidth}\centering
        \includegraphics[width=\textwidth]{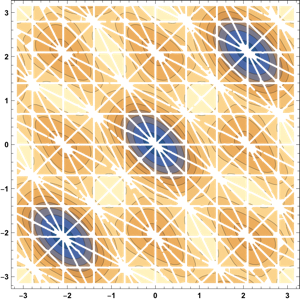}
        \caption{$G_2$ $N_a=3$}
        \label{fig:G2NA3}
    \end{subfigure}\vspace{0.2cm}\\
    \caption{Contour plot of $\text{Im}\mathcal{F}|_{\Delta = -i \pi}$ for rank-2 SYM with $N_a=2,3$ adjoint chiral multiplets.}
    \label{fig:rank2msym}
\end{figure}

\paragraph{$\CN=4$ SYM and two adjoint $\CN=1$ theory}
Our next example is $\mathcal{N}=4$ SYM with different gauge groups. This is the gauge theory with $N_a = 3$ adjoint chiral multiplets. The R-charge of the chiral multiplet is 
\begin{align}
    R &= 1-\frac{1}{N_a}.
\end{align}
Inspecting the real and imaginary value of  
\begingroup
\allowdisplaybreaks
\begin{align}
    \mathcal{F} = 
        \sum_{s = \pm}  s\sum_{\rho \in \mathbf{adj}}\Big(&
        \text{Li}_3(-e^{s\Delta + i s \rho (\vec{\alpha})} ) - N_a 
        \text{Li}_3(-e^{s(1/N_a)\Delta  +i  s \rho(\vec\alpha)})\Big),
\end{align}
\endgroup
we find that  \ref{itm:cond1}~and~\ref{itm:cond2} hold at $N_a = 2,3$. This implies that $\alpha_1 = \cdots = \alpha_{|G|} = 0$ is the most dominant saddle point of the asymptotic integral \eqref{eq:sci-integral-cardy}. See Figures~\ref{fig:rank1msym}~and~\ref{fig:rank2msym} for the plots of $\text{Im}(\mathcal{F})|_{\Delta = -i\pi}$ for the rank-1 and rank-2 gauge theories.

\begin{figure}[t!]
    \centering
    \begin{subfigure}[b]{0.3\textwidth}\centering
        \includegraphics[width=\textwidth]{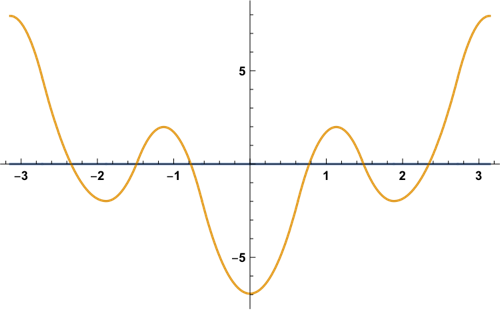}
        \caption{$SU(2)$ ISS model}
        \label{fig:ISS}
    \end{subfigure}\hspace{0.5cm}
    \begin{subfigure}[b]{0.3\textwidth}\centering
        \includegraphics[width=\textwidth]{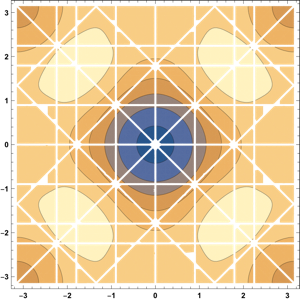}
        \caption{$SO(5)$ BCI model}
        \label{fig:BCI}
    \end{subfigure}\hspace{0.5cm}
    \begin{subfigure}[b]{0.3\textwidth}\centering
        \includegraphics[width=\textwidth]{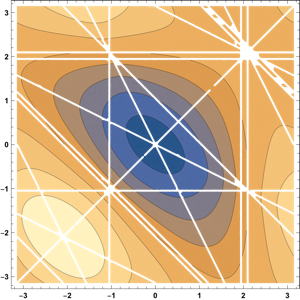}
        \caption{$SU(3)$ Pouliot model}
        \label{fig:MPo}
    \end{subfigure}
    \caption{$\text{Im}\mathcal{F}|_{\Delta = -i \pi}$ for $SU(2)$ ISS model, $SO(5)$ BCI model, and $SU(3)$ Pouliot model.}
    \label{fig:plotF2}\vspace{0.5cm}
\end{figure}

\paragraph{ISS model} The ISS model \cite{Intriligator:1994rx} is an $SU(2)$ gauge theory with one spin-$\frac{3}{2}$ chiral multiplet, which flows to an interacting fixed point in IR.\footnote{See \cite{Intriligator:2005if, Vartanov:2010xj} on the discussion of its IR phase.} The R-charge of the chiral multiplet is $R = 3/5$. In Figure~\ref{fig:ISS}, we draw
the real and imaginary value of 
\begin{align}
    \mathcal{F} = 
        \sum_{s,\sigma = \pm}  s\left(
        \text{Li}_3(-e^{s\Delta + i s  \cdot 2\sigma\alpha}) ) - 
        \text{Li}_3(-e^{s(2/5)\Delta  +i  s \cdot \sigma \alpha}) - \text{Li}_3(-e^{s(2/5)\Delta  +i  s \cdot 3\sigma \alpha})\right)
\end{align}
for $\alpha \in (-\pi, \pi)$. This shows that  \ref{itm:cond1}~and~\ref{itm:cond2} are true, therefore $\alpha = 0$ is the most dominant saddle point of the asymptotic integral \eqref{eq:sci-integral-cardy}.

\paragraph{BCI model} The BCI model \cite{Brodie:1998vv} is an $SO(N)$ gauge theory with a chiral multiplet in the rank-2 symmetric, traceless representation. The R-charge of the chiral multiplet is
\begin{align}
    R = \frac{4}{N+2}.
\end{align}
They are asymptotically free for $N\geq 5$ and flow to an interacting IR fixed point. 
\begin{align}
    \mathcal{F} = 
        \sum_{s = \pm}  s\left( \sum_{\rho \in \mathbf{adj}} \text{Li}_3(-e^{s\Delta + i s \rho (\vec{\alpha})} ) 
        - \sum_{\lambda \in \mathbf{sym}}\text{Li}_3(-e^{s(\frac{N-2}{N+2})\Delta  +i  s \lambda(\vec\alpha)})   \right).
\end{align}
We find that \ref{itm:cond1}~and~\ref{itm:cond2} holds for $5 \leq N \leq 21$. So $\alpha_1 = \cdots = \alpha_{|G|} = 0$ is the most dominant saddle point of \eqref{eq:sci-integral-cardy}. See Figure \ref{fig:BCI} as the contour plot of $\text{Im}(\mathcal{F})|_{\Delta = -i\pi}$ for the $SO(5)$ model.

\paragraph{Magnetic Pouliot theory} This model is an $SU(N)$ gauge theory ($3\leq N \leq 10$) with one symmetric and $(N+4)$ anti-fundamental chiral multiplets, plus a  meson for the $SU(N+4)$ flavor symmetry. It is dual to $Spin(7)$ gauge theory with $(N+4)$ spinor chiral multiplets \cite{Pouliot:1995zc}. The R-charge assignment of the chiral multiplets is
\begin{align}
    R_s = \frac{2}{N}, \qquad R_{af} = \frac{5}{N+4} - \frac{1}{N}, \qquad R_{s} = 2 - \frac{10}{N+4}.
\end{align}
Ignoring the contribution of gauge singlets, we study the real and imaginary values of 
\begin{align}
\begin{split}
    \mathcal{F} = 
        \sum_{s = \pm}  s\Big(& \sum_{\rho \in \mathbf{adj}} \text{Li}_3(-e^{s\Delta + i s \rho (\vec{\alpha})} ) -  \sum_{\lambda \in \mathbf{sym}}\text{Li}_3(-e^{s(1-2/N)\Delta  +i  s \lambda(\vec\alpha)}) \\[-0.1cm] 
         & \qquad \qquad - (N+4)\sum_{\lambda \in \overline{\mathbf{fnd}}}\text{Li}_3(-e^{s(\frac{N^2+4}{N(N+4)})\Delta  +i  s \lambda(\vec\alpha)})   \Big).
\end{split}
\end{align}
Again we find that \ref{itm:cond1}~and~\ref{itm:cond2} hold, thus $\alpha_1 = \cdots = \alpha_{|G|} = 0$ being the most dominant saddle point of \eqref{eq:sci-integral-cardy}. Figure~\ref{fig:MPo} is the contour plot of $\text{Im}(\mathcal{F})|_{\Delta = -i\pi}$ for the $SU(3)$ model.

\begin{figure}[t!]
    \centering
    \begin{subfigure}[b]{0.23\textwidth}\centering
        \includegraphics[width=\textwidth]{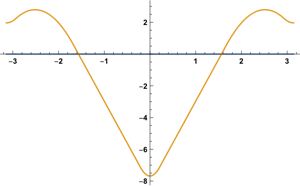}
        \caption{$(A_1, A_2)$ AD theory}
        \label{fig:ADA1A2}
    \end{subfigure}\hspace{0.2cm}
    \begin{subfigure}[b]{0.23\textwidth}\centering
        \includegraphics[width=\textwidth]{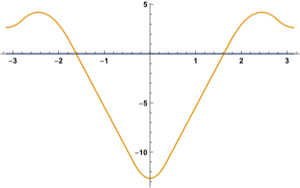}
        \caption{$(A_1, A_3)$ AD theory}
        \label{fig:ADA1A3}
    \end{subfigure}\hspace{0.2cm}
    \begin{subfigure}[b]{0.23\textwidth}\centering
        \includegraphics[width=\textwidth]{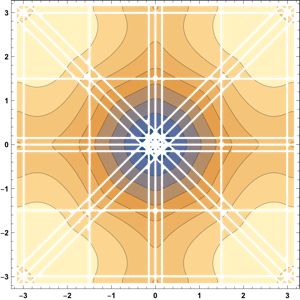}
        \caption{$(A_1, A_4)$ AD theory}
        \label{fig:ADA1A4}
    \end{subfigure}\hspace{0.2cm}
    \begin{subfigure}[b]{0.23\textwidth}\centering
        \includegraphics[width=\textwidth]{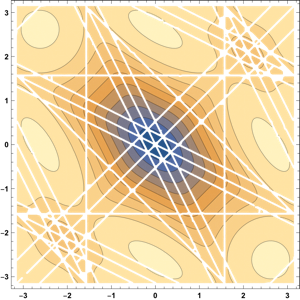}
        \caption{$(A_1, A_5)$ AD theory}
        \label{fig:ADA1A5}
    \end{subfigure}\\\vspace{0.2cm}
    \caption{$\text{Im}\mathcal{F}|_{\Delta = -i \pi}$ for $(A_1, A_N)$ Agyres-Douglas theory with $N \leq 5$.}
    \label{fig:AD}
\end{figure}

\paragraph{Argyres-Douglas theory} One can also study the $\CN=2$ Argyres-Douglas (AD) theories using their $\CN=1$ Lagrangian descriptions \cite{Maruyoshi:2016tqk, Maruyoshi:2016aim, Agarwal:2016pjo, Agarwal:2017roi, Benvenuti:2017bpg, Fluder:2017oxm, Maruyoshi:2018nod}. The $(A_1, A_{2N-1})$ AD theory can be described by the $SU(N)$ gauge theory with 1 fundamental, 1 anti-fundamental, and 1 adjoint chiral multiplets, plus a number of gauge singlets. Their $R$-charges are assigned as follows:
\begin{align}
    R_f  = R_{af} = 1+ (\epsilon-1)N, \ \  R_{a} = 1-\epsilon, 
\end{align}
where $\epsilon = \frac{3N+1}{3N+3}$. Ignoring the contribution of gauge singlets, 
\begin{align}
    \label{eq:ad1}
    \begin{split}
    \mathcal{F} = 
        \sum_{s = \pm}  s\Big(& \sum_{\rho \in \mathbf{adj}} \text{Li}_3(-e^{s\Delta + i s \rho (\vec{\alpha})} ) -  \sum_{\lambda \in \mathbf{fnd}}\text{Li}_3(-e^{s(1-\epsilon)N\Delta  +i  s \lambda(\vec\alpha)}) \\
        - &    \sum_{\lambda \in \overline{\mathbf{fnd}}}\text{Li}_3(-e^{s(1-\epsilon)N\Delta  +i  s \lambda(\vec\alpha)}) - \sum_{\lambda \in {\mathbf{adj}}}\text{Li}_3(-e^{s\epsilon\Delta  +i  s \lambda(\vec\alpha)})   \Big).
        \end{split}
\end{align}
Similarly, the $(A_1, A_{2N})$-type AD theory is described by the $Sp(N)$ gauge theory with two fundamental and one adjoint chiral multiplet, plus a number of gauge singlets. Their $R$-charges are
\begin{align}
    R_{f_1}  = \frac{1+\epsilon}{2}, \quad R_{f_2}  = (2N+\tfrac{3}{2})\epsilon - (2N + \tfrac{1}{2}), \quad R_{a} = 1-\epsilon, 
\end{align}
where $\epsilon = \frac{6N+7}{6N+9}$. Correspondingly, we consider
\begin{align}
    \label{eq:ad2}
    \begin{split}
    \mathcal{F} = 
        \sum_{s = \pm}  s\Big(& \sum_{\rho \in \mathbf{adj}} \text{Li}_3(-e^{s\Delta + i s \rho (\vec{\alpha})} ) -  \sum_{\lambda \in \mathbf{fnd}}\text{Li}_3(-e^{s((1-\epsilon)/2)\Delta  +i  s \lambda(\vec\alpha)}) \\
        & -  \sum_{\lambda \in \mathbf{fnd}}\text{Li}_3(-e^{s(2N+\frac{3}{2})(1-\epsilon)\Delta  +i  s \lambda(\vec\alpha)})   - \sum_{\lambda \in {\mathbf{adj}}}\text{Li}_3(-e^{s\epsilon\Delta  +i  s \lambda(\vec\alpha)})  \Big).
       \end{split}
\end{align}
We checked \ref{itm:cond1}~and~\ref{itm:cond2} hold with \eqref{eq:ad1}~and~\eqref{eq:ad2} for $N \leq 10$. So the most dominant saddle point of the holonomy integral \eqref{eq:sci-integral-cardy} is again at the origin. The plots of $\text{Im}(\mathcal{F})|_{\Delta = -i\pi}$ for $N=1,2$ theories are given in Figure~\ref{fig:AD}. 

We have also studied the $\CN=1$ deformed version of the AD theory (which has the smallest value of the central charge $a$ among the known 4d SCFTs \cite{Xie:2016hny, Buican:2016hnq, Maruyoshi:2018nod}) and found that the most dominant saddle is at the origin. 

\paragraph{SU(2)${}^3$ theory with trifundamentals} 
Let us consider $SU(2)^3$ gauge theory coupled via a pair of trifundamental chiral multiplets. If we add chiral multiplets in the adjoint of each $SU(2)$, this theory becomes $\CN=2$ class $\CS$ theory realized by wrapping 2 M5-branes on a genus 2 Riemann surface \cite{Gaiotto:2009we}. If we do not have the adjoint chiral multiplets, this belongs to $\CN=1$ class $\CS$ theory with the normal bundles of degree $(1, 1)$ \cite{Bah:2012dg}. The trifundamentals of $\CN=1$ and $\CN=2$ theory $R$-charge $1/2$ and $2/3$ respectively. Therefore, the central charges are
\begin{align}
\begin{split}
 a_{\CN=1} = \frac{15}{8}, \quad c_{\CN=1} = \frac{29}{16} , \quad \frac{a_{\CN=1}}{c_{\CN=1}} = \frac{30}{29} , \\
 a_{\CN=2} = \frac{53}{24}, \quad c_{\CN=2} = \frac{13}{6}  , \quad  \frac{a_{\CN=2}}{c_{\CN=2}} = \frac{53}{52} .  
\end{split}
\end{align}
Notice that $a/c>1$ for both cases. Quite generally, the class $\CS$ theories corresponding to higher genus ($g\ge 2$) Riemann surface with no puncture exhibits $a/c>1$. 

\begin{figure}[t!]
    \centering
    \begin{subfigure}[b]{0.3\textwidth}\centering
        \includegraphics[width=\textwidth]{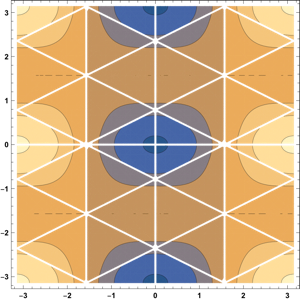}
        \caption{$\CN=1$ theory}
    \end{subfigure}\hspace{0.5cm}
    \begin{subfigure}[b]{0.3\textwidth}\centering
        \includegraphics[width=\textwidth]{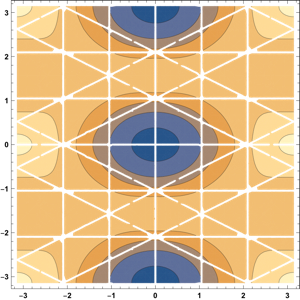}
        \caption{$\CN=2$ theory}
    \end{subfigure}
\\\vspace{0.2cm}
    \caption{$\text{Im}\mathcal{F}|_{\Delta = -i \pi}$ for the $SU(2)^3$ theories coupled via trifundamentals}
    \label{fig:trifund}
\end{figure}

We obtain
\begin{align}
 \CF = \sum_{s = \pm} s \left(\sum_{\rho = \pm 2,0, m = 1, 2, 3} \text{Li}_3 (-e^{s(\Delta + i \rho \alpha_m) })
   - 2 \sum_{w_{1, 2, 3} = \pm} \text{Li}_3(-e^{s(\half \Delta + i w_m \alpha_m)}) \right) \, 
\end{align}
for the $\CN=1$ theory and 
\begin{align}
\begin{split}
 \CF = \sum_{s = \pm} s \left( \sum_{\substack{\rho = \pm 2,0, \\m = 1, 2, 3}} \left( \text{Li}_3 (-e^{s(\Delta + i \rho \alpha_m) }) - \text{Li}_3 (-e^{s(\frac{1}{3}\Delta + i \rho \alpha_m) }) \right) 
 - 2 \sum_{w_{1, 2, 3} = \pm} \text{Li}_3(-e^{s(\frac{1}{3} \Delta + i w_m \alpha_m)}) \right) \, 
\end{split}
\end{align}
for the $\CN=2$ theory. We plot the subspace of $\CF$ with $\a_2 = \a_3$ in Figure \ref{fig:trifund}. We find that the most dominant saddle is again at the origin. 

\section{Asymptotic entropy} \label{sec:entropy}

Given the superconformal index $\mathcal{I}$, the microstate degeneracy $\Omega(J_1, J_2, R)$ can be obtained by taking an inverse Laplace transformation on $\mathcal{I}$. However, if we consider the asymptotic degeneracy at large angular momenta, i.e., $J_1 \sim J_2 \gg a, c$, it suffices to take the Legendre transformation on the Cardy free energy \eqref{eq:asymCardyF}. So let us extremize the entropy function\footnote{This form of the entropy function was conjectured in Appendix A of \cite{Hosseini:2018dob}.}
\begin{align}
    \label{eq:entropy-ftn}
    S(\Delta, \omega_{1,2}; R, J_{1,2}) = \frac{8(5a-3c)}{27\omega_1 \omega_2}\Delta^3 + \frac{8\pi^2(a-c)}{3\omega_1\omega_2}\Delta + R \Delta + J_1 \omega_1 + J_2 \omega_2,
\end{align}
under the constraint $2\Delta - \omega_1 - \omega_2 = -2\pi i$.
One should keep in mind that the Cardy free energy \eqref{eq:asymCardyF} can be trusted only up to the $\mathcal{O}(\omega^{-2})$ order.
Following the interpretation of \cite{Choi:2018hmj, Choi:2018vbz, Choi:2019miv}, we take the real part of the extremized $S$ as the asymptotic entropy of our index. Note that $\textrm{Re}(S)$ is still a priori a lower bound for the true entropy.

Extremizing $S(\Delta, \omega_{1,2}; R, J_{1,2})$ in terms of $\omega_1$ and $\omega_2$ yields the following two equations:
\begin{align}
    \label{eq:Jr/2}
    \begin{split}
    J_1 + \frac{R}{2} &= -\frac{16 i \pi ^3 (3c-2a)}{27 \omega _1^2 \omega _2} + \mathcal{O}(\omega^{-2})\\
    J_2 + \frac{R}{2} &= -\frac{16 i \pi ^3 (3c-2a)}{27 \omega _1 \omega _2^2} + \mathcal{O}(\omega^{-2})
    \end{split}
\end{align}
Subtracting these two equations, we find
\begin{align}
    \label{eq:JD}
    J_1 - J_2 = - \frac{16 i \,   (3c-2a)\pi ^3}{27 \omega _1 \omega _2}  \left(\frac{1}{\omega _1}-\frac{1}{\omega _2}\right) + \mathcal{O}(\omega^{-2}).
\end{align}
It is complicated to solve these equations generally, so we simplify our calculus by setting the two angular momenta to be equal, i.e., $J \equiv J_1 = J_2$ and $\omega \equiv \omega_1 = \omega_2$. Then \eqref{eq:JD} becomes void. We denote the real and imaginary part of the chemical potential $\omega$ by $\omega_R$ and $\omega_I$, i.e., $\omega_+ = \omega_R + i\omega_I$.
Since $J$ and $R$ must be real-valued, the imaginary part of \eqref{eq:Jr/2} becomes
\begin{align}
    \label{eq:extreme-sol}
    0 \approx  \pi (3c-2a)  \omega_R (\omega _R^2-3 \omega_I^2).
\end{align}
Demanding $\omega_R$ to be real-valued, we identify three different solutions of \eqref{eq:extreme-sol}.
\begin{align}
    \label{eq:omegar-sol}
    \omega_R = 0, \ \pm \sqrt{3} \w_I . 
\end{align}
Inserting these solutions back to \eqref{eq:Jr/2}, we find
\begin{align}
    J + \frac{R}{2} \approx 
    \begin{dcases}
        \frac{2 \pi ^3 (3 c-2 a)}{27 \omega_I^3} &\ \text{for the two solutions $\omega_R = \pm \sqrt{3} \omega_I$,}\\
       -\frac{16\pi ^3 (3 c- 2 a)}{27 \omega_I^3} &\ \text{for the solution $\omega_R = 0$.}
    \end{dcases}
\end{align}
The BPS states captured by the superconformal index $\mathcal{I}$ should carry $(J +  \frac{R }{2}) > 0$ as we discussed in section \ref{sec:index}. Thus the real part $\omega_R$ of the chemical potential should also be positive. The only solution among \eqref{eq:omegar-sol} which satisfies both requirements is
\begin{align}
    \label{eq:omega-sol-chosen}
    \omega_R \simeq \sqrt{3} \omega_I \qquad \text{for}\qquad   \omega_I >0  \ . 
\end{align}
This solution makes $\text{Re}(\log{\CI})$ to be
\begin{align}
    \label{eq:re-logz}
    \text{Re}(\log{\CI}) = \frac{2\pi^2 (3c-2a)}{9 \sqrt{3} \omega_I^2} + \CO(\omega_I^{-1}) = \frac{8\pi^2 (3c-2a) }{9\sqrt{3} \beta^2} + \CO(\omega_I^{-1}) \ ,  
\end{align}
where $\omega = \beta e^{\pi i /6}$. 

The extremized entropy $S$ is generally complex-valued. As a consistency check, we consider its real part $\text{Re}(S)$ and check if the solution \eqref{eq:omega-sol-chosen} makes $\text{Re}(S) > 0$. In fact,
\begin{align}
    \label{eq:re-s}
    \text{Re}(S) \approx 
        + \frac{2 \pi^3  (3 c-2a)}{3
   \sqrt{3}  \ \omega_I^2}  >0
\end{align}
always, thanks to the Hofman-Maldacena bound $\frac{1}{2} < \frac{a}{c} < \frac{3}{2}$ for an interacting $\mathcal{N}=1$ SCFT. Expressing the entropy in terms of the angular momentum $J$, we obtain
\begin{align}
    \label{eq:re-s-J}
    \text{Re}(S) = +2^{1/3}\,3^{1/2}\,(3c-2a)^{1/3}\,\pi \cdot  J^{2/3}+ \mathcal{O}(J^{1/3}) , 
\end{align}
which is positive as long as $a/c < 3/2$.

\subsection{Free chiral/vector theories}
It was noticed in \cite{Choi:2018hmj, Choi:2018vbz, Benini:2018ywd} that the complexified chemical potentials are crucial to obstruct the boson/fermion cancellation in the computation of the entropy at large angular momenta. Especially for $\mathcal{N}=4$ SYM,  the boson/fermion cancelation was maximally obstructed at the optimal value of chemical potentials, determined by the Legendre transformation.
At least in the large $N$ and strong-coupling limit, this entropy from $\CI$ saturates the upper bound, which is the true entropy given by the Bekenstein-Hawking entropy of dual black holes that counts BPS states without $(-1)^F$.
However, to illustrate that $\text{Re}(S)$ from our index is in general only a lower bound of the true entropy $\text{Re}(S_\text{true}) = \log{\Omega_\text{true}}$, here we compute the true BPS degeneracy of the free system and compare it with \eqref{eq:re-s}.

The partition function (as opposed to the Witten index) of a free QFT can be evaluated by counting the BPS operators satisfying $\delta_- = 0$. The chemical potentials are no longer subject to the index constraint. For a free chiral (with $R$-charge $2/3$) and vector multiplet,
\begin{align}
    \mathcal{Z}_{\mathcal{X}} &= \exp\left(\sum_{m=1}^\infty\frac{1}{m}\frac{e^{-m\beta}t^{-2m/3} + (-1)^{m+1} e^{-3m\beta/2}t^{m/3}}{(1-e^{-m\beta}p^m)(1-e^{-m\beta}q^m)}\right) , \\
    \mathcal{Z}_{V} &= \exp\left(\sum_{m=1}^\infty\frac{1}{m}\frac{ e^{-2m \beta}(pq)^m +(-1)^{m+1}t^{m} \left(e^{-3m\beta/2}\chi_2(p^mq^{-m}) - e^{-5m\beta/2}(pq)^{m/2}\right) }{(1-e^{-m\beta}p^m)(1-e^{-m\beta}q^m)}\right).\nn
\end{align}
Let us set $\Delta = 0$. In our asymptotic limit $\beta/r \ll |\omega_{1,2}| \ll 1$, they become
\begin{align}
    \log\mathcal{Z}_{\mathcal{X}} = \log\mathcal{Z}_{V} = \frac{\text{Li}_3(1) - \text{Li}_3(-1)}{\omega_1\omega_2} + \mathcal{O}(\omega^{-1}) =\frac{7\zeta(3)}{4\omega_1\omega_2} + \mathcal{O}(\omega^{-1}).
\end{align}
Taking the Legendre transformation, we find the entropy of a free vector/chiral multiplet as
\begin{align}
    S^{\rm true}(J_1,J_2) = \frac{7\zeta(3)}{4\omega_1\omega_2} + J_1 \omega_1 + J_2 \omega_2 \, \bigg|_{\omega_{i} = \omega_{i}^*}\ \simeq \ 4.467\, (J_1 J_2)^{1/3}.
\end{align}
On the other hand, the asymptotic entropy \eqref{eq:re-s-J} captured in the index is 
\begin{align}
    \text{Re}(S) = \begin{dcases}2.995\,J^{2/3} & \text{for a free chiral multiplet, }\\
        0 & \text{for a free vector multiplet}.
    \end{dcases}
\end{align}
at the equal momenta $J_1 = J_2$. Since $\text{Re}(S) < S^{\rm true}$ for these cases, we conclude that the extremized entropy $\text{Re}(S)$ from the index does \emph{not} always exhibit the maximum degeneracy.
The above calculation means that turning on interactions can lift some of the BPS states that the index does not count. It is still possible (but hard to prove or disprove) that our $\text{Re}(S)$ may equal to the asymptotic $S^{\rm true}$ for the interacting SCFTs as was in the case of $\CN=4$ SYM theory. It will be interesting to understand this issue better, perhaps by studying more exotic BPS black holes in AdS$_5$ beyond the known ones.

\subsection{Holographic SCFTs and AdS${}_5$ black holes}

Here, let us apply our asymptotic entropy formula \eqref{eq:re-s} to holographic SCFTs. It is natural to expect that this accounts for the Bekenstein-Hawking entropy of various BPS black holes in asymptotic AdS${}_5$. For a precision check of this correspondence, here we once again perform the Legendre transformation of the Cardy free energy at $\omega_1 \neq \omega_2$ with non-trivial flavor chemical potentials. 

Our main example is a family of $\mathcal{N}=1$ superconformal quiver theory dual to type IIB supergravity on AdS${}_5 \times Y^{p,p}$ \cite{Benvenuti:2004dy}. 
This gauge theory is obtained from $N$ D3-branes probing $\mathbb{C}^3/\mathbb{Z}_{2p}$ orbifold. 
It has $2p$ gauge groups and $4p$ bifundamental chiral multiplets. In addition to the $U(1)_R$ symmetry, there are flavor symmetries $U(1)_B$, $U(1)_F$, and $SU(2)_l$. All the bifundamental chiral multiplets are divided into three different species, denoted as $U$, $V$, $Y$. 
For each type of multiplet, the number of fields and representation under $U(1)_R \times U(1)_B \times U(1)_F \times SU(2)_L$ are summarized in the following table:
\begin{align*}
    \begin{tabular}{c|c|cccc}
        & Number & $U(1)_R$ & $U(1)_B$ & $U(1)_F$ & $SU(2)_L$\\\hline
        $U$&$p$ & $2/3$ &$-p$ & $0$ & $\bf 2$\\
        $V$&$p$ & $2/3$ & $p$ & $\frac{1}{2}$ & $\bf 2$\\
        $Y$&$2p$ & $2/3$ & $0$ & $-\frac{1}{2}$ & $\bf 1$
    \end{tabular}
\end{align*}
We refer to \cite{Benvenuti:2004dy} for a detailed description of $Y^{p,q}$ quiver gauge theory, for all $0\leq q\leq p$. 
The $Y^{p,q}$ superconformal index in the large $N$ limit agrees is shown to agree with the BPS graviton index on AdS${}_5 \times Y^{p,q}$ \cite{Nakayama:2006ur, Gadde:2010en, Eager:2012hx}. 

Let us introduce two flavor chemical potentials $\Delta_B$, $\Delta_F$, conjugate to $U(1)_B$, $U(1)_F$ at zero $SU(2)_L$ charge. This is the case in which the BPS black hole solutions are known in AdS$_5 \times S^5/Z_{2p}$ \cite{Gutowski:2004ez, Gutowski:2004yv, Chong:2005hr, Kunduri:2006ek} via a $U(1)^3$ Kaluza-Klein reduction \cite{Cvetic:1999xp}. 
The non-vanishing anomaly coefficients (in the large $N$ limit) are
\begin{align}
    \text{Tr}(R^3)&=  \frac{16p}{9}N^2, & \text{Tr}(RF^2) &= -\frac{p}{3}N^2, & \text{Tr}(RB^2) &= -\frac{4 p^3}{3}N^2, \\
     \text{Tr}(RBF) &= -\frac{p^2}{3}N^2, & \text{Tr}(FB^2) &= p^3N^2, & \text{Tr}{(BF^2)} &= \frac{p^2}{2}N^2.
\end{align}
We arrange the $U(1)_R \times U(1)_B \times U(1)_F$ chemical potentials into the following combinations:
\begin{align}
    \Delta_1 \equiv\frac{2}{3} \Delta + \frac{i}{2} x_F,\quad \Delta_2 \equiv \frac{2  }{3}\Delta + i p x _B, \quad \Delta_3 \equiv \frac{2  }{3}\Delta-  \frac{i}{2}x _F - i p x _B.
\end{align}
Here, $x_B$, $x_F$ are the chemical potentials associated to $U(1)_B$ and $U(1)_F$ respectively. 
They are subject to the index constraint  $\Delta_1 + \Delta_2+\Delta_3  -\omega_1-\omega_2 = -2\pi i$. 
The corresponding entropy function $S\big(\Delta_{1,2,3}, \omega_{1,2}; R, F, B, J_{1,2}\big)$ is given by (with $\tilde{B} \equiv \frac{1}{2p}B$)
\begin{align}
    S = p N^2\cdot \frac{\Delta_1 \Delta_2 \Delta_3}{\omega_1\omega_2} + (R-F)\Delta_1 + (R-\tilde{B})\Delta_2 + (R+F+\tilde{B})\Delta_3 + \sum_{i=1}^2 J_i\omega_i.
\end{align}
Now we extremize $S$ under the constraint  $\Delta_1 + \Delta_2+\Delta_3  -\omega_1-\omega_2 = -2\pi i$. This is exactly the same type as was studied in \cite{Hosseini:2017mds}. Repeating the same procedure as in \cite{Choi:2018hmj}, we find the following cubic equation in $S$:
\begin{align}
    \label{eq:cubic-s}
    \Big(\frac{S}{2\pi i}- R+F\Big)\Big(\frac{S}{2\pi i}- R+\tilde{B}\Big)\Big(\frac{S}{2\pi i}- R-F-\tilde{B}\Big) = pN^2\Big(\frac{S}{2\pi i} +  J_1\Big)\Big(\frac{S}{2\pi i} +  J_2\Big)
\end{align}
This equation has 3 complex solutions in general. Any physically relevant solution that represents a black hole should satisfy $\text{Re}(S)/ N^{2} > 0$ with all the $U(1)^3$ charges and two angular momenta are of $\mathcal{O}(N^2)$. Let us focus on the special case of $\text{Im}(S)=0$. 

In fact, BPS black holes in AdS$_5 \times S^5 / Z_{2p}$ are known in this circumstance \cite{Gutowski:2004ez, Gutowski:2004yv, Chong:2005hr, Kunduri:2006ek}.
Dividing the above equation \eqref{eq:cubic-s} into the real and imaginary parts, we obtain
\begin{align}
\begin{split}
    0 &= (3R+pN^2)\, S^2 -4 \pi ^2 \left((R-F)(R-\tilde{B})(R+F+\tilde{B}) + pN^2 J_1 J_2\right) \ ,  \\
    0 &= S^3 -4\pi^2 S \left(3 R^2-F^2-\tilde{B}^2-\tilde{B} F -pN^2(J_1+J_2)\right).
\end{split}
\end{align}
Solving for $S$, we get
\begin{align}
    \label{eq:entropy-ypp}
    \begin{split}
    \frac{S}{2\pi} &=\sqrt{\frac{(R-F)(R-\tilde{B})(R+F+\tilde{B}) + pN^2 J_1J_2}{3R+pN^2}} \\
    &=  \sqrt {3 R^2-F^2-\tilde{B}^2-\tilde{B} F -pN^2(J_1+J_2)} \ .
    \end{split}
\end{align}
Compatibility of these two expressions implies the charge relation of the AdS${}_5$ black hole. Especially at large angular momenta $J \gg N^2$, the charge relation implies $R,F,B \simeq \mathcal{O}(J^{2/3})$. Once we insert the charge relation back to \eqref{eq:entropy-ypp}, we obtain the entropy as $S \simeq \sqrt{3} \,(pN^2)^{1/3} J^{2/3} + \mathcal{O}(J^{1/3})$, which agrees with \eqref{eq:re-s-J} at $a=c = pN^2/2$.





\acknowledgments
We thank Abhijit Gadde, Kimyeong Lee, Sungjay Lee, June Nahmgoong, Wenbin Yan and Piljin Yi for helpful discussions. 
This work is supported in part by the National Research Foundation of Korea (NRF) Grants 2018R1A2B6004914 (SK) and 2017R1D1A1B06034369 (JS).
The work of JS is supported by the Junior Research Group Program at the APCTP through the Science and Technology Promotion Fund, Lottery Fund of the Korean Government, Gyeongsangbuk-do, and Pohang City.
The work of JS is also supported by the National Research Foundation of Korea (NRF) grant NRF-2020R1C1C1007591 and the Settlement Research Grant for the new faculty provided by Korea Advanced Institute for Science and Technology (KAIST). The work of JK is supported by the NSF grant PHY-1911298.

\bibliographystyle{JHEP}
\bibliography{ref}

\end{document}